\documentclass[apj, 8pt]{emulateapj-rtx4}
\usepackage{pdflscape}
\usepackage{graphicx}
\usepackage{scalefnt}
\usepackage[bookmarks=false]{hyperref}
\renewcommand\thefootnote{\arabic{footnote}}

\slugcomment{accepted for publication in ApJ}

\shortauthors{T.Shibuya et al.}

\begin{document}

\title{What is the physical origin of strong Ly$\alpha$ emission? \\ II. Gas Kinematics and Distribution of Ly$\alpha$ Emitters  \altaffilmark{\ddag}}

%% AUTHOR
%%%%%%%%%%%%%%%%%%%%%%%%%%%%%%%%%%%%%%%%%%%%%%%%

%\author{Takatoshi SHIBUYA\altaffilmark{1,2} et al.}
\author{Takatoshi Shibuya\altaffilmark{1,2}, Masami Ouchi\altaffilmark{1,3}, Kimihiko Nakajima\altaffilmark{3,4}, Takuya Hashimoto\altaffilmark{3}, Yoshiaki Ono\altaffilmark{1,3}, Michael Rauch\altaffilmark{5}, Jean-Rene Gauthier\altaffilmark{6}, Kazuhiro Shimasaku\altaffilmark{3,7}, Ryosuke Goto\altaffilmark{3}, Masao Mori\altaffilmark{2}, and Masayuki Umemura\altaffilmark{2}.}
\email{shibyatk\_at\_icrr.u-tokyo.ac.jp}

\altaffiltext{1}{Institute for Cosmic Ray Research, The University of Tokyo, 5-1-5 Kashiwanoha, Kashiwa, Chiba 277-8582, Japan}
\altaffiltext{2}{Center for Computational Sciences, The University of Tsukuba, 1-1-1 Tennodai, Tsukuba, Ibaraki 305-8577 Japan}
\altaffiltext{3}{Department of Astronomy, Graduate School of Science, The University of Tokyo, Tokyo 113-0033, Japan}
\altaffiltext{4}{Institute for the Physics and Mathematics of the Universe (IPMU), TODIAS, The University of Tokyo, 5-1-5 Kashiwanoha, Kashiwa, Chiba 277-8583, Japan}
\altaffiltext{5}{Observatories of the Carnegie Institution of Washington, 813 Santa Barbara Street, Pasadena, CA 91101, USA}
\altaffiltext{6}{Cahill Center for Astronomy and Astrophysics, California Institute of Technology, MS 249-17, Pasadena, CA 91125, USA}
\altaffiltext{7}{Research Center for the Early Universe, Graduate School of Science, The University of Tokyo, Tokyo 113-0033, Japan}

\altaffiltext{\ddag}{Based on data obtained with the Subaru Telescope operated by the National Astronomical Observatory of Japan.}

%% ABSTRACT
%%%%%%%%%%%%%%%%%%%%%%%%%%%%%%%%%%%%%%%%%%%%%%%%

\begin{abstract}

We present a statistical study of velocities of Ly$\alpha$, interstellar (IS) absorption, and nebular lines and gas covering fraction for Ly$\alpha$ emitters (LAEs) at $z\simeq 2$. We make a sample of 22 LAEs with a large Ly$\alpha$ equivalent width (EW) of $\gtrsim 50$\,\AA\ based on our deep Keck/LRIS observations, in conjunction with spectroscopic data from the Subaru/FMOS  program and the literature. We estimate the average velocity offset of Ly$\alpha$ from a systemic redshift determined with nebular lines to be $\Delta v_{\rm Ly\alpha}=234\pm 9$ km s$^{-1}$. Using a Kolmogorv-Smirnov test, we confirm the previous claim of Hashimoto et al. (2013) that the average $\Delta v_{\rm Ly\alpha}$ of LAEs is smaller than that of LBGs. Our LRIS data successfully identify blue-shifted multiple IS absorption lines in the UV continua of four LAEs on an individual basis. The average velocity offset of IS absorption lines from a systemic redshift is $\Delta v_{\rm IS}=204 \pm 27$ km s$^{-1}$, indicating LAE's gas outflow with a velocity comparable to typical LBGs. Thus, the ratio, $ R^{\rm Ly\alpha}_{\rm IS} \equiv \Delta v_{\rm Ly\alpha}/\Delta v_{\rm IS}$ of LAEs, is around unity, suggestive of low impacts on Ly$\alpha$ transmission by resonant scattering of neutral hydrogen in the IS medium. We find an anti-correlation between Ly$\alpha$ EW and the covering fraction, $f_c$, estimated from the depth of absorption lines, where $f_c$ is an indicator of average neutral hydrogen column density, $N_{\rm HI}$. The results of our study support the idea that $N_{\rm HI}$ is a key quantity determining Ly$\alpha$ emissivity.
 
\end{abstract}


%% INTRODUCTION: 
%%%%%%%%%%%%%%%%%%%%%%%%%%%%%%%%%%%%%%%%%%%%%%%%

\section{INTRODUCTION}

Ly$\alpha$ Emitters (LAEs) are an important population of high-$z$ star-forming galaxies in the context of galaxy formation.  LAEs at $z=2-7$ and beyond $z=7$ are found by narrow-band (NB) imaging observations based on an NB excess resulting from their prominent Ly$\alpha$ emission \citep[e.g., ][]{2010ApJ...711..928C,2007ApJ...667...79G,2012ApJ...744..110C,2008ApJS..176..301O,2008ApJ...677...12O,2010ApJ...723..869O,2010ApJ...725..394H,2007ApJ...660.1023F,2011ApJ...734..119K,2006ApJ...648....7K,2012ApJ...752..114S}. Observational studies on a morphology and spectral energy distribution (SED) of LAEs reveal that such a galaxy is typically young, compact, less-massive, less-dusty than other high-$z$ galaxy populations, and a possible progenitor of Milky Way mass galaxies \citep[e.g., ][]{2011ApJ...743....9G,2011ApJ...733..114G,2010MNRAS.402.1580O,2007ApJ...671..278G,2011ApJ...740...71D,2008ApJ...681..856R,2012MNRAS.424.1672D}. Additionally, LAEs are used to measure the neutral hydrogen fraction at the reionizing epoch, because Ly$\alpha$ photons are absorbed by intergalactic medium (IGM). 

Ly$\alpha$ emitting mechanism is not fully understood due to the highly-complex radiative transfer of Ly$\alpha$ in the interstellar medium (ISM). Many theoretical models have predicted that the neutral gas and/or dust distributions surrounding central ionizing sources are closely linked to the Ly$\alpha$ emissivity \citep[e.g., ][]{1991ApJ...370L..85N, 2008ApJ...678..655F,2013ApJ...766..124L,2009ApJ...704.1640L,2007ApJ...657L..69L,2013arXiv1302.7042D,2013arXiv1308.1405Z,2010ApJ...716..574Z,2012arXiv1209.5842Y}. Thus, resonant scattering in the neutral ISM can significantly attenuate the Ly$\alpha$ emission. 

Ly$\alpha$ emissivity may not only depend on the spatial ISM distribution, but on the gas kinematics as well. The large-scale galactic outflows driven by starbursts or active galactic nuclei could allow Ly$\alpha$ photons to emerge at wavelengths where the Gunn-Peterson opacity is reduced, and consequently enhance the Ly$\alpha$ emissivity, particularly in the high-$z$ Universe \citep[e.g., ][]{2010MNRAS.408..352D}. The outflow may also blow out the Ly$\alpha$ absorbing ISM. The gas kinematics of LAEs has been evaluated from the Ly$\alpha$ velocity offset ($\Delta v_{{\rm Ly\alpha}}$) with respect to the systemic redshift ($z_{\rm sys}$) traced by nebular emission lines (e.g, H$\alpha$, $[$O {\sc iii}$]$) from their H {\sc ii} regions. In the past few years, deep NIR spectroscopic studies have detected nebular emission lines from $\sim10$ LAEs at $z=2-3$, and measured their $\Delta v_{{\rm Ly\alpha}}$ \citep{2011ApJ...730..136M, 2013ApJ...765...70H,2013A&A...551A..93G,2011ApJ...729..140F,2013ApJ...775...99C}. The Ly$\alpha$ emission lines for these LAEs are redshifted from their $z_{\rm sys}$ by a $\Delta v_{{\rm Ly\alpha}}$ of $200-300$ km s$^{-1}$. \citet{2013ApJ...765...70H} find an anti-correlation between Ly$\alpha$ equivalent width (EW) and $\Delta v_{{\rm Ly\alpha}}$ in a compilation of LAE and LBG samples. This result is in contrast to a simple picture where Ly$\alpha$ photons more easily escape in the presence of a galactic outflow.

However, the Ly$\alpha$ velocity offset is thought to increase with both resonant scattering in H {\sc i} gas clouds as well as galactic outflow velocity \citep[e.g., ][]{2006A&A...460..397V,2008A&A...491...89V}. The anti-correlation could result from a difference in H {\sc i} column density ($N_{\rm HI}$) rather than outflowing velocity. The gas kinematics can be investigated more directly from the velocity offset between interstellar (IS) absorption lines of the rest-frame UV continuum and $z_{\rm sys}$ (IS velocity offset; $\Delta v_{{\rm IS}}$). The IS velocity offset traces the speed of outflowing gas clouds, and may help to distinguish the two effects on $\Delta v_{{\rm Ly\alpha}}$. 

For UV-continuum selected galaxies, the $\Delta v_{{\rm IS}}$ has been measured for $>100$ objects \citep[e.g.][]{2001ApJ...554..981P,2012MNRAS.427.1973C,2012ApJ...745...33K,2013ApJ...777...67S,2010ApJ...717..289S}. \citet{2010ApJ...717..289S} find that LBGs have an average of $\langle\Delta v_{{\rm IS}}\rangle = -164$ km s$^{-1}$ in their sample of 89 LBGs at $z\sim3$. This statistical study indicates ubiquitousness of galactic outflow in LBGs. However, there have been no NB-selected galaxies with a $\Delta v_{{\rm IS}}$ measurement to date except for a stacked UV spectrum in \citet{2013ApJ...765...70H}. This is because it is difficult to estimate $\Delta v_{{\rm IS}}$ for individual LAEs, especially for galaxies with a large Ly$\alpha$ EW of $\gtrsim 50$\,\AA\, due to their faint UV-continuum emission, while $\Delta v_{{\rm IS}}$ are measured for some UV-selected galaxies with EW(Ly$\alpha) \sim50$\,\AA\, \citep[e.g., ][]{2010ApJ...719.1168E}. A statistical investigation of Ly$\alpha$ kinematics for LAEs could shed light on the physical origin of the anti-correlation and the underlying Ly$\alpha$ emitting mechanism.

%% EXPLANATION OF THIS PAPER
%%%%%%%%%%%%%%%%%%%%%%%%%%%%%%%%%%%%%%%%%%%%%%%%

This is the second paper in the series exploring the Ly$\alpha$ emitting mechanisms\footnote{The first paper presents a study on LAE structures \citep{2014ApJ...785...64S}.}. In this paper, we present the results of our optical and NIR spectroscopy for a large sample of $z=2.2$ LAEs with Keck/LRIS and Subaru/FMOS to verify possible differences of $\Delta v_{{\rm Ly\alpha}}$ and $\Delta v_{{\rm IS}}$ between LAEs and LBGs. These spectroscopic observations are in an extension of the project of \citet{2013ApJ...765...70H} aiming to confirm the anti-correlation between Ly$\alpha$ EW and $\Delta v_{{\rm Ly\alpha}}$. The organization of this paper is as follows. In Section \ref{sec_targets}, we describe the details of the LAEs targeted for our spectroscopy. Next, we show our optical and NIR spectroscopic observations in Section \ref{sec_observation}. We present methods to reduce the spectra, and to measure kinematic quantities such as $\Delta v_{{\rm Ly\alpha}}$ and $\Delta v_{{\rm IS}}$ in \ref{sec_reduction}. We perform SED fitting to derive physical properties in Section \ref{sec_sed_fitting}. We compare kinematic properties between LAEs and LBGs in Section \ref{sec_results}, and discuss physical origins of possible differences in these quantities in Section \ref{sec_discussion}. In the last section Section \ref{sec_conclusion}, we summarize our findings. 

Throughout this paper, we adopt the concordance cosmology with $(\Omega_m, \Omega_\Lambda, h)=(0.3, 0.7, 0.7)$, \citep{2011ApJS..192...18K}. All magnitudes are given in the AB system \citep{1983ApJ...266..713O}.

%% LAE SAMPLE
%%%%%%%%%%%%%%%%%%%%%%%%%%%%%%%%%%%%%%%%%%%%%%%%
\section{TARGETS for SPECTROSCOPY}\label{sec_targets}

Our targets for optical and NIR spectroscopy are $z=2.2$ LAEs selected by observations of the Subaru/Suprime-Cam \citep{2002PASJ...54..833M} equipped with the narrow-band (NB) filter, NB387 ($\lambda_c = 3870$ \AA\, and FWHM $= 94$ \AA) \citep{2012ApJ...745...12N,2013ApJ...769....3N}. The details of observations and selection for LAEs are given in these papers, but we provide a brief description as follows. The Suprime-Cam observations have been carried out for LAEs at $z=2.2$ with NB387 in a total area of $\sim1.5$ square degrees. Based on the color selection of $B-NB387$ and $u^*-NB387$, the Suprime-Cam observations have located $619$, $919$, $747$, $950$, and $168$ LAEs in the Cosmic Evolution Survey (COSMOS) \citep{2007ApJS..172....1S}, the Subaru/{\it XMM-Newton} Deep Survey (SXDS) \citep{2008ApJS..176....1F}, the Chandra Deep Field South (CDFS) \citep{2001ApJ...551..624G}, the {\it Hubble} Deep Field North (HDFN) \citep{2004ApJ...600L..93G}, and the SSA22 \citep[e.g., ][]{2000ApJ...532..170S} fields, respectively. In the above five fields, a total of $\sim3400$ LAEs have been selected down to a Ly$\alpha$ EW of $20-30$ \AA\, in rest-frame (Nakajima et al. in prep.). This large sample size enables us to study statistically various properties of high-$z$ LAEs, such as structural properties \citep{2014ApJ...785...64S} and the statistics of Ly$\alpha$ halos \citep{2014arXiv1403.0732M}.

%% OBSERVATION
%%%%%%%%%%%%%%%%%%%%%%%%%%%%%%%%%%%%%%%%%%%%%%%%
\section{OBSERVATION}\label{sec_observation}

%%%%%%%%%%%%%%%%%%%%%%%%%%%%%%%%%%%%%%%%%%%%%%%%
\subsection{Optical Spectroscopy for Ly$\alpha$ and UV Continuum Emission}\label{subsec_optical_spec}

We have carried out optical spectroscopy for our $z=2.2$ LAE sample with the Low Resolution Imaging Spectrometer \citep[LRIS; ][]{1995PASP..107..375O,2004ApJ...604..534S} on the Keck I telescope in order to detect their redshifted Ly$\alpha$ emission lines. We used six multi-object slit (MOS) masks for LAEs selected in the NB387 imaging observations in the COSMOS, HDFN, HUDF, SSA22, and SXDS fields. The mask for the objects in the HUDF includes two LAEs whose nebular emission lines were detected in the 3D-{\it HST} survey (H. Atek et al. in preparation).  The total number of LAEs observed with these LRIS masks is 83. The observations were conducted on March 19-21 and November 14-15, 2012 (UST) with seeing sizes of $0.\!\!^{\prime\prime}7$-$1.\!\!^{\prime\prime}6$. Spectrophotometric standard stars were observed on each night for flux calibrations. The spectral resolution is $R\sim1000$. The number of observed LAEs, grisms, central wavelength and observing time in each slit-masks are summarized in Table \ref{table_observation_log}. 
%The spectral resolution is $R\sim1000$, which corresponds to $\sim200$ km s$^{-1}$ at $z=2.2$. 

\begin{deluxetable*}{ccccccc}
\tablecolumns{7}
\tablewidth{0pt}
\setlength{\tabcolsep}{0.30cm} 
\tabletypesize{\scriptsize}
\tablecaption{Summary of the Keck/LRIS Observations}
\tablehead{  \colhead{Slit Mask} & \colhead{$n_{\rm LAE}/n_{\rm obj}$} & \colhead{Grating/$\lambda_c$} & \colhead{$t_1$} & \colhead{$n_{\rm frame}$} &  \colhead{$T_{\rm exp}$} & \colhead{Date of Observations} \\ 
\colhead{}& \colhead{}& \colhead{[\AA]}& \colhead{[s]} &  \colhead{} &  \colhead{[s]} &  \colhead{} \\
\colhead{(1)}& \colhead{(2)}& \colhead{(3)}& \colhead{(4)} &  \colhead{(5)} &  \colhead{(6)} &  \colhead{(7)}} 

\startdata
    COSMOS & $14/16$ & $600/4000$ & $3000$ & $8$ & $24000$ & 2012 March 19-21  \\ 
    HDFN1 & $18/22$ & $600/4000$ & $3000$ & $6$ &  $18000$ & 2012 March 20  \\
    HDFN2 & $18/20$ & $1200/3400$ & $2800-3000$ & $6$ &  $17800$ & 2012 March 19-20  \\
    COSMOS3B & $18/22$ & $600/4000$ & $3000$ & $3$ &  $9000$ & 2012 November 15 \\
    HUDF\_maB & $10/31$ & $400/3400$ & $2758-3000$ & $2$ &  $5758$ & 2012 November 14 \\
    SXDS495B & $10/30$ & $600/4000$ & $2136-3000$ & $14$ &  $40854$ & 2012 November 14-15 
\enddata
\tablecomments{Columns: (1) Slit mask. (2) Number of objects included in the slit mask. (3). Grating and the central wavelength. (4) Exposure time of one frame. (5) Number of exposure. (6) Exposure time. (7) Date of observations.}
\label{table_observation_log}
\end{deluxetable*}

%%%%%%%%%%%%%%%%%%%%%%%%%%%%%%%%%%%%%%%%%%%%%%%%
\subsection{Near-Infrared Spectroscopy for Nebular Emission}\label{subsec_nir_spec}

To calculate systemic redshifts of our LAEs from their nebular emission lines, we use NIR spectroscopic data obtained from observations with the Fiber Multi Object Spectrograph \citep[FMOS; ][]{2010PASJ...62.1135K} on the Subaru telescope on December 22, 23, and 24, 2012 (UST). All of LAEs in the SXDS and COSMOS fields are observed with J and H-band filters of FMOS. Details of the FMOS observation and reduction are shown in Nakajima et al. in prep. The systemic redshifts for objects were derived by simultaneously fitting to H$\beta$ and $[$O {\sc iii}$]\lambda4958, 5007$ emission lines by using their vacuum wavelengths in rest-frame.

%% REDUCTION
%%%%%%%%%%%%%%%%%%%%%%%%%%%%%%%%%%%%%%%%%%%%%%%%
\section{Spectroscopic Data}\label{sec_reduction}

%%%%%%%%%%%%%%%%%%%%%%%%%%%%%%%%%%%%%%%%%%%%%%%%
\subsection{Reduction of LRIS Spectra}\label{subsec_reduction}

Our LRIS spectra in each MOS mask are reduced with the public Low-Redux ({\tt XIDL}) pipeline\footnote{http://www.ucolick.org/\,$\tilde{ }$\,xavier/LowRedux/}, for longslit and multi-slit data from the spectrographs on the Keck, Gemini, MMT, and Lick telescopes. We reduce the spectra of LAEs with this software in the following manner. First, we create flats, calibrate wavelengths with the arc data, and reject sources illuminated by the cosmic ray injections for 2-D spectra in the MOS masks. Next, we automatically identify emission lines and continua, and trace them in each slit in individual one-frame masks. After the source identification, we subtract the sky background, and correct for the distortion of the 2-D MOS mask images using sky lines. According to the information on the source identifications, we extract 1-D spectra from each slit in individual mask images. Finally, we stack the extracted 1-D spectra. 

The public {\tt XIDL} software extracts 1-D spectra from each one-exposure frame {\it before} combining these 2-D mask images. This process makes it difficult to detect faint emission lines and continua that are undetectable in individual one-exposure images. Then, we additionally search for faint emission lines from stacked 2-D images by visual inspection {\it after} combining one-exposure frames. 

\begin{figure}[t!]
  \begin{center}
    \includegraphics[width=80mm]{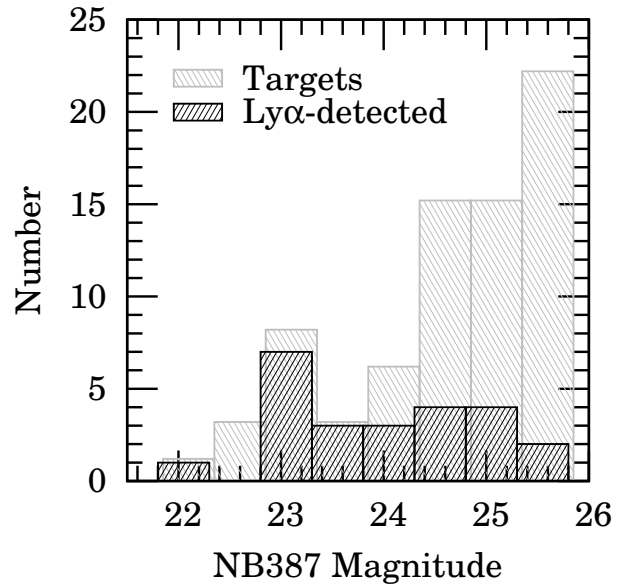}
  \end{center}
  \caption[IS absorption lines of LAEs]{{\footnotesize Success rate in identifying a Ly$\alpha$ emission line in the LRIS spectroscopy as a function of NB387 magnitude. The gray histogram shows the number of targeted LAEs. The black histogram indicate the number of LAEs with a Ly$\alpha$ detection. The histogram of targeted LAEs is slightly shifted for clarity.}}
  \label{fig_hist_success_rate}
\end{figure}

In total, the Ly$\alpha$ emission lines are detected from 26 objects in the LRIS spectroscopy. Figure \ref{fig_hist_success_rate} shows the spectroscopic success rate in the detection of Ly$\alpha$ emission. The success rate is $\sim70$\% for bright objects with NB387$\lesssim24.5$. However, low detection and/or selection completeness at NB387$\gtrsim24.5$ reduces largely the success rate ($\sim20$\%). The photometric and spectroscopic properties of these Ly$\alpha$-detected objects are listed in Table \ref{table_keck_lae}. Among these LRIS spectra, we identify eight LAEs with detections of Ly$\alpha$ and nebular emission lines excluding AGN-like objects. 

%%%%%%%%%%%%%%%%%%%%%%%%%%%%%%%%%%%%%%%%%%%%%%%%
\subsection{Measurement of Ly$\alpha$ Velocity Offset}\label{subsec_measure_lya_voffset}

We measure the Ly$\alpha$ velocity offset for the eight LAEs with detections of Ly$\alpha$ and nebular lines:  

\begin{equation}\label{eq_voffset}
	\Delta v_{\rm Ly\alpha} = c \frac{z_{\rm  Ly\alpha} - z_{\rm sys}}{1+z_{\rm sys}}, 
\end{equation}

where $c$, $z_{\rm  Ly\alpha}$, and $z_{\rm sys}$, are the speed of light, and Ly$\alpha$ and systemic redshifts, respectively. The systemic redshift is determined from nebular emission lines obtained with FMOS. 

Prior to the measurement of $\Delta v_{\rm Ly\alpha}$, we measure the wavelength of Ly$\alpha$ in the following line-fitting procedures. We use the peak wavelength of the best-fit asymmetric Gaussian profile for measurements of the Ly$\alpha$ wavelength. We first search automatically for an emission line in a wavelength range of $3500-4000$\,\AA\, in each spectrum. This range includes the wavelength range of the NB387 filter. Next, we fit an asymmetric Gaussian profile to the detected lines. The asymmetric Gaussian profile is expressed as 

\begin{equation}\label{eq_asymgauss}
        f(\lambda) = A \exp \left(\frac{-(\lambda-\lambda_0^{\rm asym})^2}{2 \sigma_{\rm asym}^2}\right) + f_0, 
\end{equation}

where $A$, $\lambda_0$, and $f_0$ are the amplitude, peak wavelength of the emission line, and continuum level, respectively. The asymmetric dispersion, $\sigma_{\rm asym}$, is represented by $\sigma_{\rm asym} = a_{\rm asym} (\lambda-\lambda_0^{\rm asym})+d$, where $a_{\rm asym}$ and $d$ are the asymmetric parameter and typical width of the line, respectively. An object with a positive (negative) $a_{\rm asym}$ value has a skewed line profile with a red (blue) wing. The fitting with the asymmetric Gaussian profile is efficient for Ly$\alpha$ line from high-$z$ galaxies affected by complex kinematic structure of infalling and/or outflowing gas and IGM absorption. For fitting, we use data points over the wavelength range where the flux drops to $10$\% of its peak value at the redder and bluer sides of the emission line. We use the peak flux, wavelength of the line peak, $0.4$, $1.0\times10^{-17}$, and $2.0$ as the initial parameters of $A$, $\lambda_0^{\rm asym}$, $a_{\rm asym}$, $f_0$, and $d$ for the line-fitting. The last two are typical values of our spectra. If profile fitting does not converge to the minimum in $\chi^2$, we search for the best-fit by changing the initial value of $a_{\rm asym}$. 

\begin{figure}[t!]
  \begin{center}
    \includegraphics[width=80mm]{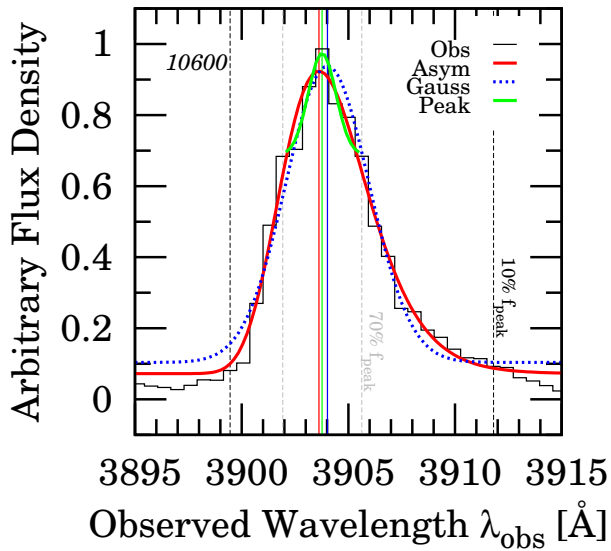}
  \end{center}
  \caption[HST images]{{\footnotesize Observed Ly$\alpha$ emission line (black) for an example LAE, 10600, and its best-fit profiles. The curves are the best-fit symmetric Gaussian profiles in the wavelength range where the flux drops to $70$ \% (green) and $10$ \% (blue) of its peak, and the asymmetric Gaussian profile (red). The vertical bold lines denote the corresponding peak wavelengths of the best-fit profiles. The peak wavelengths are 3903.77, 3904.02, and 3903.62\,\AA, respectively, with the central, symmetric Gaussian, and asymmetric Gaussian profiles. The vertical dashed lines indicate the wavelengths where the flux drops to $70$ \% (gray) and $10$ \% (black) of its peak. See details in \S \ref{subsec_measure_lya_voffset}. }}
  \label{fig_linefit}
\end{figure}

We show the best-fit asymmetric Gaussian profile for an example spectrum in Figure \ref{fig_linefit}. We also fit a {\it symmetric} Gaussian profile to the emission lines in addition to asymmetric one. For the symmetric Gaussian fitting, we adopt two wavelength ranges where the flux drops to $70$\% and $10$\% of its peak value, and denote the corresponding peak wavelengths by $\lambda_0^{\rm gauss}$ and $\lambda_0^{\rm cent}$, respectively. The fitting procedure in the former narrow range is similar as in \citet{2013ApJ...765...70H} in terms of avoiding systematic effects due to asymmetric line profile. As shown in Fig. \ref{fig_linefit}, the best-fit $\lambda_0^{\rm asym}$ is broadly equal to $\lambda_0^{\rm cent}$ for the example line. The wavelength difference is $\sim+0.1$\,\AA ($\sim+10$ km s$^{-1}$ at $z=2.2$). In contrast, $\lambda_0^{\rm gauss}$ differs from $\lambda_0^{\rm asym}$ by $\sim+0.4$\,\AA\, which corresponds to a velocity difference of $\sim+30$ km s$^{-1}$ at $z=2.2$. This is likely to be caused by the sharp drop on the blue side and the extended red tail which cannot be fit well with symmetric profiles.

\begin{figure}[t!]
  \begin{center}
    \includegraphics[width=90mm]{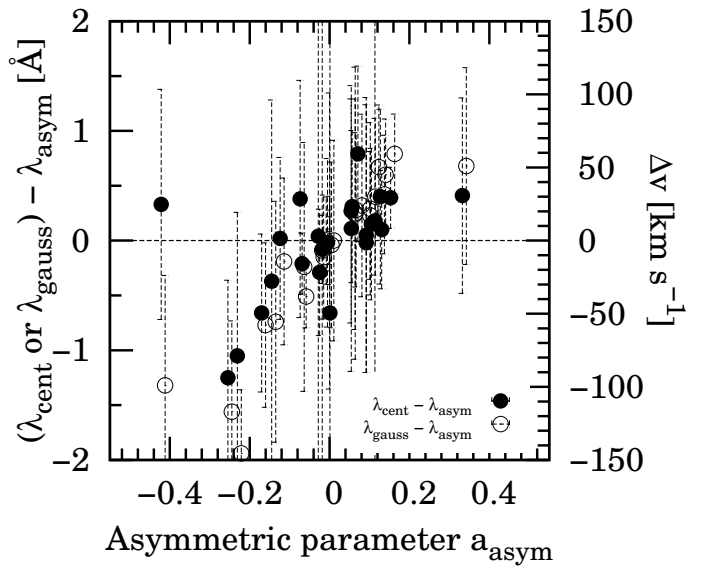}
  \end{center}
  \caption[Lya Wavelength]{{\footnotesize Differences between peak wavelengths of the best-fit central (filled circles) or symmetric Gaussian (open circles) profiles and ones of asymmetric Gaussian profile for Ly$\alpha$-detected objects. Right vertical axis indicates the corresponding velocity offset. See details in Section \ref{subsec_measure_lya_voffset}.}}
  \label{fig_lya_wavelength}
\end{figure}

This trend is more clearly shown in Figure \ref{fig_lya_wavelength} which exhibits the wavelength difference of $\lambda_0^{\rm gauss}$ and $\lambda_0^{\rm cent}$ from $\lambda_0^{\rm asym}$ as a function of the asymmetric parameter, $a_{\rm asym}$. The wavelengths of individual profiles are in good agreement for almost symmetric lines with $a_{\rm asym}\sim0$. Even for moderately-asymmetric profile with $|a_{\rm asym}|\lesssim0.2$, $\lambda_0^{\rm cent}$ tends to correct for systematic effects of skewed lines compared to $\lambda_0^{\rm gauss}$. However, both of $\lambda_0^{\rm gauss}$ and $\lambda_0^{\rm cent}$ are redshifted (blueshifted) from $\lambda_0^{\rm asym}$ by $\sim0.5-1.0$\,\AA\, for highly-asymmetric lines with $a_{\rm asym}\sim+0.4$ ($-0.4$).

\begin{figure}[t]
  \begin{center}
    \includegraphics[width=80mm]{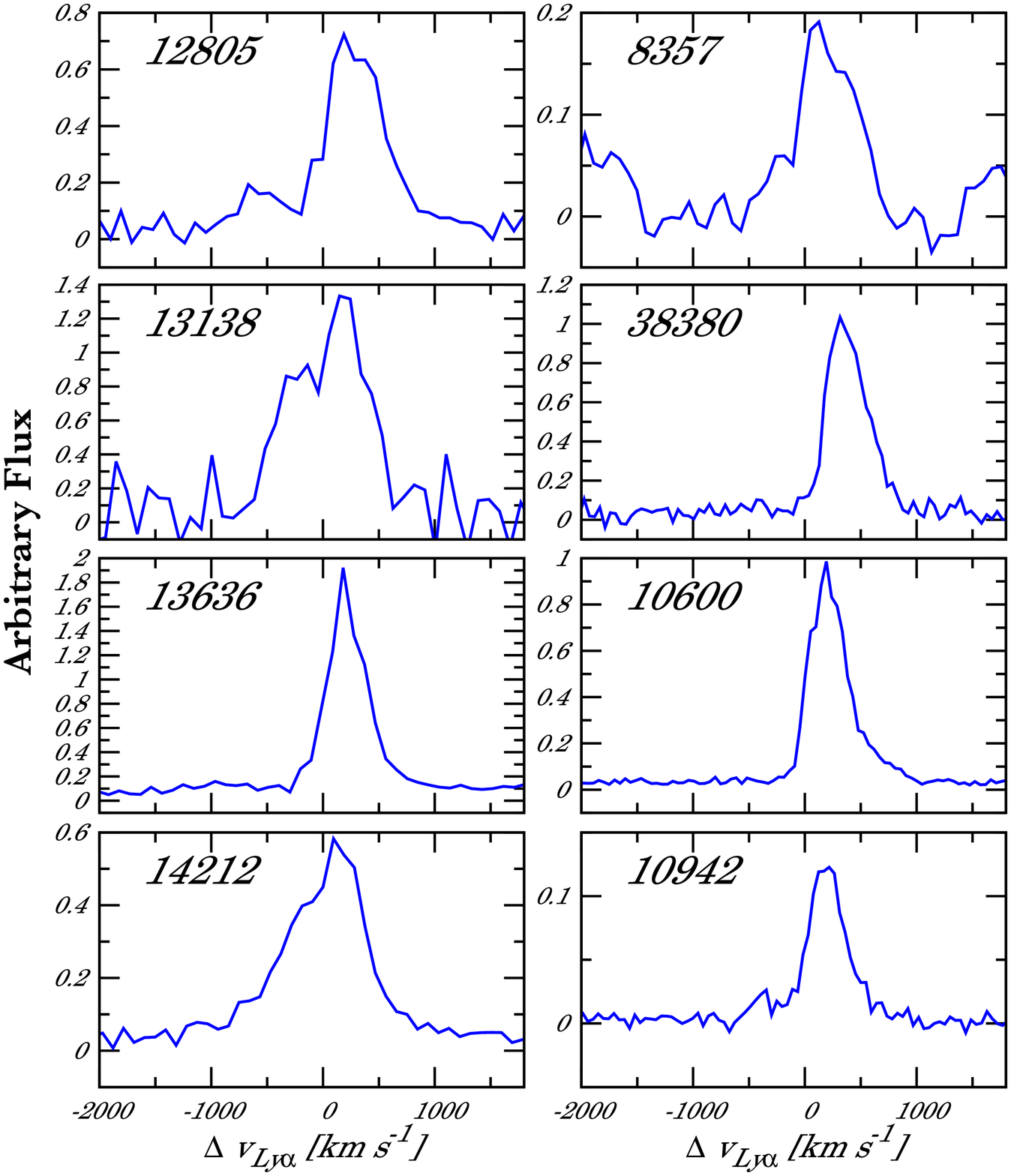}
  \end{center}
  \caption[IS absorption lines of LAEs]{{\footnotesize Ly$\alpha$ emission lines as a function of velocity for the eight LAEs with detections of nebular lines.}}
  \label{fig_plottable}
\end{figure}

After correcting for the heliocentric motion of Earth for the redshifts of Ly$\alpha$ and nebular lines\footnote{http://fuse.pha.jhu.edu/support/tools/vlsr.html}, we calculate $\Delta v_{\rm Ly\alpha}$ following Equation \ref{eq_voffset}. Table \ref{table_keck_lae} lists the $z_{\rm  Ly\alpha}$, $z_{\rm  sys}$, and $\Delta v_{\rm Ly\alpha}$ for the 26 Ly$\alpha$-detected objects observed with LRIS. Figure \ref{fig_plottable} present Ly$\alpha$ spectra as a function of velocity for LAEs with detections of nebular emission lines. In Table \ref{table_imacs_lae}, we also list these quantities of the four LAEs with detections of Ly$\alpha$ and nebular lines obtained by previous Magellan/IMACS observations \citep{2012ApJ...745...12N}. Almost all objects observed with LRIS have a $\Delta v_{\rm Ly\alpha}$ of $\sim200$ km s$^{-1}$ which is consistent with values in previous studies \citep[e.g., ][]{2013ApJ...765...70H}. The values of $\Delta v_{\rm Ly\alpha}$ in the IMACS sample are calculated to be smaller than the LRIS results. This could be caused by large uncertainties due to the IMACS spectroscopy with a lower spectral resolution than LRIS.

\begin{turnpage}
\begin{deluxetable*}{clccccccccccccc}
\setlength{\tabcolsep}{0.15mm} 
\tabletypesize{\scriptsize}
\tablecaption{Summary of the Ly$\alpha$-detected Objects in the LRIS Spectroscopy}
\tablehead{\colhead{Slit Mask} & \colhead{Object} & \colhead{R.A.} & \colhead{Decl.} & \colhead{\it{U}} & \colhead{NB387} & \colhead{\it{B}} & \colhead{$\lambda_{\rm obs}$} & \colhead{$z_{\rm Ly\alpha}$} & \colhead{$f(Ly\alpha)$} & \colhead{$L(Ly\alpha)$} & \colhead{EW(Ly$\alpha$)} & \colhead{$z_{\rm sys}$} & \colhead{$\Delta v_{\rm Ly\alpha}$} & \colhead{$F_{\rm blue} / F_{\rm tot}$} \\ 
\colhead{}& \colhead{}& \colhead{}& \colhead{}& \colhead{}& \colhead{}& \colhead{}& \colhead{[\AA]} & \colhead{} &  \colhead{[$10^{-17}$erg/s/cm$^2$]} &  \colhead{[$10^{42}$erg/s]} & \colhead{[\AA]} & \colhead{} & \colhead{[km s$^{-1}$]} & \colhead{} \\
\colhead{(1)}& \colhead{(2)}& \colhead{(3)}& \colhead{(4)} &  \colhead{(5)} &  \colhead{(6)} &  \colhead{(7)} & \colhead{(8)} & \colhead{(9)} & \colhead{(10)} &  \colhead{(11)} &  \colhead{(12)} & \colhead{(13)} & \colhead{(14)} & \colhead{(15)}}

\startdata
COSMOS & 12027 & 149.9343976 & +2.1285326 & 24.2 & 23.4 & 24.5& $3878.72\pm0.24$ & 2.1906 & $5.6\pm0.3$ & $2.0\pm0.1$ & $73.69^{+6.91}_{-6.44}$ & \nodata & \nodata & \nodata \\ 
		& 12805\tablenotemark{b} & 150.0637013 & +2.1354116 & 23.7 & 23.3 & 23.8 & $3843.27\pm0.65$ & 2.16144 & $7.4\pm0.7$ & $2.6\pm0.3$ & $33.73^{+5.98}_{-5.52}$ & 2.15872 & $258\pm51$ & $0.24$\\ 
		& 13138 & 150.0108585 & +2.1401388 & 24.9 & 24.6 & 25.0 & $3866.73\pm0.89$ & 2.18074 & $1.2\pm0.2$ & $0.43\pm0.07$ & $40.36^{+9.64}_{-8.44}$ & 2.17921 & $144\pm69$ & $0.40$\\ 
		& 13636\tablenotemark{bd} & 149.9974498 & +2.1439906 & 23.9 & 23.0 & 24.1 & $3844.68\pm1.30$ & 2.1626 & $9.6\pm0.5$ & $3.3\pm0.2$ & $86.80^{+8.31}_{-7.76}$ & 2.16052 & $197\pm102$\tablenotemark{f} & $0.13$ \\ 
		& 14212\tablenotemark{b} & 149.9585714 & +2.1482830 & 24.0 & 23.3 & 24.0 & $3879.99\pm0.52$ & 2.19165 & $6.7\pm0.6$ & $2.4\pm0.2$ & $54.98^{+5.56}_{-5.18}$ & 2.18955 & $188\pm40$ & $0.23$\\ 
		& 08357\tablenotemark{a} & 149.9961405 & +2.0921070 & 24.8 & 24.4 & 24.9 & $3868.79\pm0.86$ & 2.18243 & $1.4\pm0.4$ & $0.50\pm0.14$ & $46.68^{+8.60}_{-7.69}$ & 2.18044 & $205\pm66$ & $0.13$\\
		& 13820\tablenotemark{a} & 149.9554179 & +2.1470628 & 25.6 & 25.1 & 25.9 & $3820.20\pm1.01$ & 2.14246 & $2.1\pm0.2$ & $0.72\pm0.07$ & $98.86^{+31.30}_{-26.40}$ & \nodata & \nodata & \nodata \\  
		& 14135\tablenotemark{a} & 149.9770609 & +2.1508410 & 27.0 & 25.9 & 26.8 & $3893.06\pm1.08$ & 2.2024 & $0.62\pm0.1$ & $2.3\pm0.4$ & $100.02^{+30.70}_{-24.26}$ & \nodata & \nodata \\ \hline 
HDFN1 	& 18325\tablenotemark{c} & 189.0973399 & +62.1014179 & 22.9 & 21.6 & 23.2 & $3858.78\pm0.25$ & 2.1742 & $30.0\pm0.9$  & $10.6\pm0.3$ & $151.61^{+3.52}_{-3.45}$ & \nodata & \nodata & \nodata \\ 
		& 20042\tablenotemark{a} & 189.0293966 & +62.1176510 & 25.4 & 24.7 & 25.6 & $3864.87\pm2.12$ & 2.17921 & $1.5\pm0.4$ & $0.53\pm0.14$ & $108.42^{+12.48}_{-11.36}$ & \nodata & \nodata & \nodata \\\hline
HDFN2	& 31902 & 189.3127706 & +62.2091548 & 25.4 & 23.9 & 24.9 & $3865.31\pm0.31$ & 2.17957 & $1.6\pm0.2$ & $0.6\pm0.08$ & $146.30^{+13.35}_{-12.31}$ & \nodata & \nodata & \nodata \\ 
		& 43408 & 189.4532215 & +62.2639356 & 26.7 & 25.4 & 26.3 & $3886.59\pm0.47$ & 2.19708 & $5.6\pm0.2$ & $2.0\pm0.1$ & $72.09^{+6.29}_{-5.89}$ & \nodata & \nodata & \nodata \\ 
		& 42659\tablenotemark{a} & 189.4575590 & +62.2917868 & 25.9 & 23.8 & 25.9 & $3882.07\pm0.61$ & 2.19336 & $1.4\pm0.1$ & $0.51\pm0.04$ & $661.41^{132.72}_{-104.76}$ & \nodata & \nodata & \nodata \\ \hline  
COSMOS3B & 38380 & 149.9205873 & +2.3844960 & 24.4 & 23.4 & 24.5 & $3909.79\pm0.47$ & 2.21616 & $6.9\pm0.7$ & $2.6\pm0.3$ & $137.19^{+14.80}_{-13.48}$ & 2.21256 & $336\pm36$ & $0.01$\\ 
		& 43982\tablenotemark{cd} & 149.9766453 & +2.4416582 & 24.3 & 23.2 & 24.6 & $3883.01\pm0.73$ & 2.19413 & $6.9\pm0.6$ & $2.6\pm0.2$ & $130.06^{+12.35}_{-11.35}$ & 2.19333 & $75\pm56$ & 0.41 \\ 
		& 46597 & 149.9415665 & +2.4688913 & 24.7 & 23.8 & 24.7 & $3857.99\pm0.74$ & 2.17355 & $3.5\pm0.6$ & $1.2\pm0.2$ & $58.75^{+7.07}_{-6.53}$ & \nodata & \nodata & \nodata \\
		& 38019\tablenotemark{a} & 149.9020418 & +2.3815371 & 25.9 & 25.0 & 25.7 & $3900.36\pm0.49$ & 2.2084 & $1.5\pm0.2$ & $5.5\pm0.7$ & $146.34^{+31.88}_{-26.40}$ & \nodata & \nodata & \nodata \\
		& 40792\tablenotemark{a} & 149.9444266 & +2.4094991 & 26.7 & 25.5 & 27.4 & $3901.43\pm0.53$ & 2.20928 & $2.3\pm0.4$ & $0.85\pm0.14$ & $394.30^{+121.34}_{-88.39}$ & \nodata & \nodata & \nodata \\
		& 41547\tablenotemark{a} & 149.9246216 & +2.4166699 & 26.0 & 24.9 & 26.6 & $3832.08\pm0.65$ & 2.15224 & $2.5\pm0.6$ & $0.86\pm0.21$ & $298.88^{+77.85}_{-61.81}$ & \nodata & \nodata & \nodata \\
		& 44993\tablenotemark{a} & 149.9744788 & +2.4530529 & 26.5 & 25.0 & 26.5 & $3907.65\pm0.80$ & 2.2144 & $2.1\pm0.4$ & $0.78\pm0.15$ & $240.61^{+53.56}_{-43.25}$ & \nodata & \nodata & \nodata \\ \hline  
HUDF\_maB & 17001\tablenotemark{e} & \nodata & \nodata & \nodata & \nodata & \nodata & $3705.44\pm0.73$ & 2.04806 & $22.0\pm4.0$ & $6.7\pm1.2$ & \nodata & \nodata & \nodata & \nodata \\
		& 31000\tablenotemark{e} & \nodata & \nodata & \nodata & \nodata & \nodata & $4969.63\pm0.47$ & 3.08798 & $3.7\pm0.6$ & $3.1\pm0.5$ & \nodata & \nodata & \nodata & \nodata \\ \hline 
%		& 18000?\tablenotemark{*} & ---  & ---  & --- & --- & 3DHST \\ 
SXDS495B & 06713 & 34.4224906 & -5.1136338 & 24.5 & 23.5 & 24.6 & $3894.18\pm1.02$ & 2.20332 & $3.9\pm0.2$ & $1.4\pm0.1$ & $118.77^{+5.24}_{-5.04}$ & \nodata & \nodata & \nodata \\ %2.14899  & $5172\pm80 ??$ & \nodata \\ 
%		& 08651 & 4051.20 & 2.33248332 & 1.7e-17 & 1.0e42 & \nodata \\
		& 10600\tablenotemark{b} & 34.4420541 & -5.0486039 & 23.7 & 23.0 & 23.6 & $3903.62\pm0.69$ & 2.21109 & $5.2\pm0.2$ & $1.9\pm0.1$ & $58.19^{+2.62}_{-2.54}$ & 2.20915 & $181\pm53$ & $0.03$ \\
		& 10942 & 34.4980945 & -5.0428800 & 25.6 & 24.2 & 25.6 & $3887.51\pm0.55$ & 2.19783 & $0.77\pm0.05$ & $0.28\pm0.02$ & $134.94^{+9.24}_{-8.68}$ & 2.19557 & $212\pm42$ & $0.12$ \\
		& 10535\tablenotemark{a} & 34.4246768 & -5.0488535 & 25.9 & 24.8 & 26.2 & $3905.50\pm0.36$ & 2.21263 & $3.4\pm0.2$ & $1.3\pm0.1$ & $223.09^{+22.04}_{-19.96}$ & \nodata &  \nodata & \nodata 
\enddata

\tablecomments{Columns: (1) Slit mask. (2) Object ID. (3) Right ascension. (4) Declination. (5)-(7) $U$, NB387, and $B$-band magnitudes. (8) Observed wavelength of Ly$\alpha$ measured by the asymmetric gaussian fitting. (9) Redshift of Ly$\alpha$ corrected for the heliocentric motion.  (10) Ly$\alpha$ flux uncorrected for slit loss in LRIS spectroscopy. (11) Ly$\alpha$ luminosity. (12) Ly$\alpha$ equivalent width estimated from the NB387 magnitudes. A Ly$\alpha$ position in the transmission curve are taken into account from the spectroscopic redshift of Ly$\alpha$. (13) Redshift of nebular emission lines corrected for the heliocentric motion (K. Nakajima et al. in preparation.). (14) Ly$\alpha$ velocity offset relative to nebular emission lines. (15) Ratio of Ly$\alpha$ flux in the bluer side relative to the systemic redshift to total Ly$\alpha$ flux. }
%\tablenotetext{??}{Nakajima-kun is checking the reliability of the systemic redshift measurement. }
\tablenotetext{a}{These objects are reduced without the Keck/LRIS public pipeline.}
\tablenotetext{b}{UV continuum-detected LAEs.}
\tablenotetext{c}{AGN-like objects.}
\tablenotetext{d}{These objects have also been observed with Magellan/MagE in \citet{2013ApJ...765...70H}.}
\tablenotetext{e}{K. Nakajima et al. in preparation.}
\tablenotetext{f}{\citet{2013ApJ...765...70H} have reported that LAE 13636 has a $\Delta v_{\rm Ly\alpha}$ of $99^{+16}_{-16}$ km s$^{-1}$. However, the H$\alpha$ line profile would have been affected by a residual of a neighboring OH line due to the low spectral resolution of their Keck-II/NIRSPEC observation ($R\sim1500$), making it difficult to determine accurately the systemic redshift. Our FMOS spectroscopy with $R\sim2200$ would securely detect nebular emission less affected by OH lines. }
\label{table_keck_lae}
\end{deluxetable*}
\end{turnpage}

Additionally, we provide a consistency check for our measurement of $z_{\rm Ly\alpha}$ by using the same object as in \citet{2013ApJ...765...70H}, COSMOS-13636. The object has been observed with both of LRIS in this work and Magellan/MagE in a previous work. The redshift of Ly$\alpha$ estimated from the LRIS spectrum ($z_{\rm Ly\alpha}^{\rm LRIS}=2.1626\pm0.00073$) is in good agreement with that of MagE ($z_{\rm Ly\alpha}^{\rm MagE}=2.16229\pm0.00008$) within a $1\sigma$ fitting error. The difference in velocity is $30\pm70$ km s$^{-1}$. The large error in $z_{\rm Ly\alpha}^{\rm LRIS}$ is likely to be due to the lower spectral resolution of LRIS ($R\sim1000$) than that of MagE ($R\sim4100$).

\begin{figure}[t]
  \begin{center}
    \includegraphics[width=90mm]{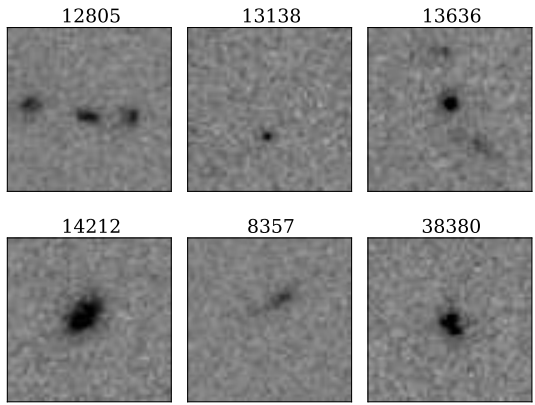}
  \end{center}
  \caption[HST images]{{\footnotesize {\it HST} $I_{814}$-band images of the LAEs with a $z_{\rm sys}$ observed with Keck/LRIS in the COSMOS field. The image size is $2^{\prime\prime}\times2^{\prime\prime}$. North is up and east is to the left. }}
  \label{fig_hst_images}
\end{figure}

Figure \ref{fig_hst_images} shows the {\it HST}/ACS $I_{814}$-band images of LAEs with a $z_{\rm sys}$ measurement in the COSMOS field. Unfortunately, the LAEs in the SXDS field are not covered by the CANDELS project. Several LAEs have multiple components, which could be mergers. The merger fraction of LAEs and its Ly$\alpha$ dependence are discussed in \citet{2014ApJ...785...64S}.

\begin{deluxetable}{cccccc}
\setlength{\tabcolsep}{0.01cm}
\tabletypesize{\scriptsize}
\tablecaption{Summary of the Ly$\alpha$-detected Objects in the IMACS Spectroscopy}
\tablehead{\colhead{Object} & \colhead{$\lambda_{\rm obs}$} & \colhead{$z_{\rm Ly\alpha}$} & \colhead{EW(Ly$\alpha$)} & \colhead{$z_{\rm sys}$} & \colhead{$\Delta v_{\rm Ly\alpha}$} \\ 
\colhead{}& \colhead{[\AA]} & \colhead{} & \colhead{[\AA]} & \colhead{} & \colhead{[km s$^{-1}$]}\\
\colhead{(1)}& \colhead{(2)}& \colhead{(3)}& \colhead{(4)} &  \colhead{(5)} &  \colhead{(6)}}

\startdata
04640 & $3865.08\pm0.37$ & 2.17938 & $164.75^{+4.63}_{-4.51}$ & 2.17822 & $110\pm138$ \\ 
		08204 & $3895.11\pm5.59$ & 2.20408 & $88.61^{+5.87}_{-5.56}$ & 2.20329 & $74\pm505$ \\ 
		09219 & $3890.71\pm6.08$ & 2.20047 & $29.62^{+2.38}_{-2.29}$ & 2.20004 & $40\pm508$ \\ 
		11135 & $3882.27\pm0.56$ & 2.19352 & $111.96^{+4.78}_{-4.60}$ & 2.19238 & $107\pm151$ 
\enddata
\tablecomments{Columns: (1) Object ID. (2) Observed wavelength of Ly$\alpha$ measured by the asymmetric gaussian fitting. (3) Redshift of Ly$\alpha$ corrected for the heliocentric motion. (4) Ly$\alpha$ equivalent width. (5) Redshift of nebular emission lines corrected for the heliocentric motion (K. Nakajima et al. in preparation.).  (6) Velocity offset of Ly$\alpha$ relative to nebular emission lines. }
\label{table_imacs_lae}
\end{deluxetable}

%%%%%%%%%%%%%%%%%%%%%%%%%%%%%%%%%%%%%%%%%%%%%%%%
\subsection{Measurement of IS Velocity Offset}\label{subsec_measure_is_voffset}

We measure the IS velocity offset of IS absorption lines for our LAEs. Due to the faintness of their UV continuum emission, it is difficult to detect IS absorption lines from high Ly$\alpha$ EW galaxies with $EW\gtrsim50$\,\AA\, in individual spectra. However, owing to the high sensitivity of Keck/LRIS, the rest-frame UV continuum emission is clearly detected from four individual LAEs, LAE 12805, 13636, and 14212 in COSMOS, and LAE 10600 in SXDS, among the 26 Ly$\alpha$-detected objects. 

\begin{figure*}[t!]
  \begin{center}
    \includegraphics[width=140mm]{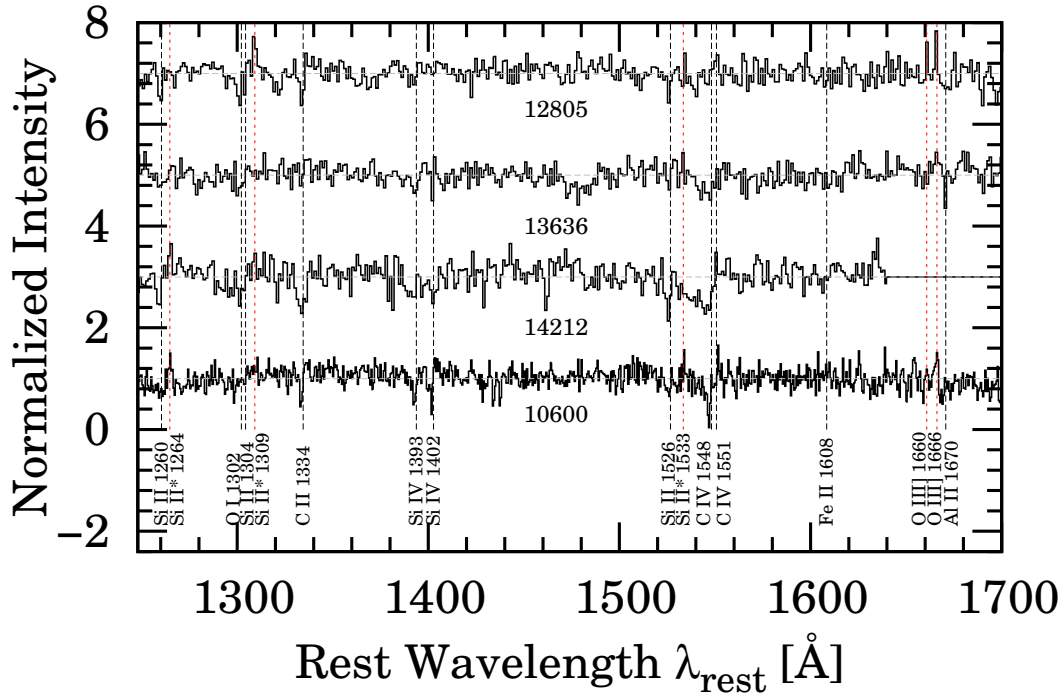}
  \end{center}
  \caption[IS absorption lines of LAEs]{{\footnotesize Normalized UV spectra of the four continuum detected LAEs. Black and red vertical dashed lines indicates wavelengths of IS absorption and emission lines, respectively. }}
  \label{fig_uvabs}
\end{figure*}

We first fit a power-law curve to the UV continuum emission in four individual objects in order to normalize the continuum level, and derive the properties of IS absorption lines. The normalized continuum emission in the rest-frame is shown in Figure \ref{fig_uvabs}. Next, we fit the symmetric Gaussian profile to each IS absorption line in a wavelength range of $\pm5$\,\AA\, around the expected line center. We summarize the best-fit peak wavelength, line depth, width, and equivalent width in Table \ref{table_abs_line}. The noise in each line is estimated from spectra at $1250-1700$\,\AA\, avoiding regions close to the absorption lines. Most absorption lines are found to be detected at the $>5\sigma$ levels except for several LIS lines. 

We calculate $\Delta v_{\rm IS}$ in a similar manner as for Ly$\alpha$ in Section \ref{subsec_lya_voffset}. Several pairs of absorption lines such as O {\sc i} $\lambda1302$-S {\sc i} $\lambda1304$, and C {\sc iv} $\lambda1548$-C {\sc iv} $\lambda1550$ are likely to be blended at the resolution of our spectroscopy. For this reason, we define the wavelengths of the line pairs as central values between the pairs. We also derive the properties of fine-structure emission lines such as Si {\sc ii}$^*$ as summarized in Table \ref{table_emi_line}. We find that the velocity offsets of these ion lines from $z_{\rm sys}$ are almost zero, indicating that the fine-structure emission lines also trace the systemic redshift of galaxies. This is because these emission lines come from nebular regions photoionized by radiation from massive stars \citep[e.g., ][]{2003ApJ...588...65S}.

%% COVERING FRACTION
%%%%%%%%%%%%%%%%%%%%%%%%%%%%%%%%%%%%%%%%%%%%%%%%
\subsection{Measurement of HI Covering Fraction}\label{subsec_measure_fc}

We estimate the covering fraction, $f_c$, of surrounding H {\sc i} gas from the depth of low ionization IS absorption lines for our four continuum-detected LAEs. If the H {\sc i} gas is distributed in a spherical shell, the depth of the lines may be related to $f_c$. The covering fraction of any ion is estimated from 

\begin{equation}\label{eq_intensity}
   \frac{I}{I_0} = 1 - f_c (1 - e^{-\tau}), 
\end{equation}

where $\tau$, $I$, and $I_0$ are optical depth of an absorption line, its residual intensity, and the continuum level, respectively. The optical depth is liked to the column density as 

\begin{equation}
   \tau = f \lambda \frac{\pi e^2}{m_e c} N = f\lambda\frac{N}{3.768\times10^{14}}, 
\end{equation}

where $f$, $\lambda$, and $N$ are the ion oscillator strength, wavelength of the absorption line in \AA, and the column density of the ion in cm$^{-2}$ (km s$^{-1}$)$^{-1}$, respectively. \citet{2013ApJ...779...52J} use Si {\sc ii} $\lambda1260$, $1304$, and $1526$ lines in order to solve the above two equations, and estimate $f_c$ for gravitationally-lensed LBGs at $z\sim4$. They find best-fit values of $N$ and $f_c$ by fitting observed the Si {\sc ii} line profiles with the intensity as a function of $N$ and $f_c$, $I(N, f_c)$, derived from the above equations. In addition to the fitting to Si {\sc ii} lines, they use several strong absorption lines, Si {\sc ii} $\lambda1260$, O {\sc i} $\lambda1302$, Si {\sc ii} $\lambda1304$, C {\sc ii} $\lambda1334$, and Si {\sc ii} $\lambda1526$ to put a lower limit on $f_c$ via 

\begin{equation}
    f_c = 1 - \frac{I}{I_0}
\end{equation}

which is a simplified case of Equation \ref{eq_intensity} when $\tau\gg1$. For our LAEs, we estimate $f_c$ in the latter method for the following reasons: (1) it is relatively difficult to fit our Si {\sc ii} line profiles with a low S/N due to the faintness of UV-continuum emission and low resolution of our spectroscopy; and (2) \citet{2013ApJ...779...52J} use mainly the $f_c$ value derived in the latter method in their discussion. We would like to compare $f_c$ for LAEs with that for LBGs in the same manner.

\begin{deluxetable*}{cccccccc}
\setlength{\tabcolsep}{0.3cm} 
\tabletypesize{\scriptsize}
\tablecaption{Absorption Line Features of the UV-continuum Detected LAEs}
\tablehead{  \colhead{Object} & \colhead{Ion} & \colhead{$\lambda_{\rm rest}$} & \colhead{$\lambda_{\rm rest}^{\rm sys}$} & \colhead{$I/I_0$} & \colhead{$\sigma$} & \colhead{EW(IS)} & \colhead{$\Delta v_{\rm IS}$} \\ 
\colhead{}& \colhead{}& \colhead{[\AA]}& \colhead{[\AA]}& \colhead{} &  \colhead{[\AA]} &  \colhead{[\AA]} &  \colhead{[km s$^{-1}$]}\\
\colhead{(1)}& \colhead{(2)}& \colhead{(3)}& \colhead{(4)} &  \colhead{(5)} &  \colhead{(6)} &  \colhead{(7)} & \colhead{(8)}}

\startdata
12805	& Si {\sc ii} & 1260.4221 & $1259.64\pm0.43$ & $-0.69\pm0.56$ & $0.67\pm0.58$ & $-1.15\pm0.21$ & $-186\pm101$ \\ 
EW(Ly$\alpha$)$=33.73$ [\AA] & O {\sc i} & 1302.1685 & $1301.02\pm0.61$ & $-0.65\pm0.33$ & $0.97\pm0.54$ & $-1.56\pm0.21$ & [$-518\pm139$] \\ 
$\Delta v_{\rm Ly\alpha}=258$ [km s$^{-1}$] & Si {\sc ii} & 1304.3702 & $1301.02\pm1.52$ & $-0.65\pm0.33$ & $0.97\pm0.54$ & $-1.56\pm0.21$ & [$-518\pm139$] \\ 
$m_B = 23.8$ & C {\sc ii} & 1334.5323 & $1333.57\pm0.72$ & $-0.58\pm0.33$ & $0.90\pm0.50$ & $-1.32\pm0.21$ & $-215\pm162$ \\
		& Si {\sc iv} & 1393.76018 & $1392.39\pm1.40$ & $-0.22\pm0.27$ & $0.64\pm0.76$ & $-0.36\pm0.21$ & $-295\pm307$ \\
		& Si {\sc ii} & 1526.70698 & $1525.91\pm0.36$ & $-0.60\pm0.17$ & $0.63\pm0.17$ & $-0.94\pm0.21$ & $-156\pm71$ \\
		& C {\sc iv} & 1548.204 & $1545.60\pm1.05$ & $-0.27\pm0.29$ & $0.70\pm0.70$ & $-0.48\pm0.21$ & [$-750\pm203$] \\
		& C {\sc iv} & 1550.781 & $1545.60\pm1.05$ & $-0.27\pm0.29$ & $0.70\pm0.70$ & $-0.48\pm0.21$ & [$-750\pm203$] \\
		& Fe {\sc ii} & 1608.45085 & $1607.75\pm1.11$ & $-0.37\pm0.48$ & $0.49\pm0.47$ & $-0.46\pm0.19$ & $-131\pm207$ \\
		& Al {\sc ii} & 1670.7886 & $1670.90\pm2.13$ & $-0.35\pm0.37$ & $1.63\pm1.9$ & $-1.43\pm0.22$ & $21\pm382$ \\ \hline
%%%%%%%%%%%%%%%%%
13636	& Si {\sc ii} & 1260.4221 & $1260.23\pm0.70$ & $-0.21\pm0.068$ & $1.76\pm0.65$ & $-0.91\pm0.19$ & $-46\pm165$ \\ 
EW(Ly$\alpha)=86.80$ [\AA] & O {\sc i} & 1302.1685 & $1301.25\pm1.20$ & $-0.33\pm0.26$ & $1.26\pm1.07$ & $-1.05\pm0.19$ & [$-465\pm280$] \\ 
$\Delta v_{\rm Ly\alpha}=197$ [km s$^{-1}$] & Si {\sc ii} & 1304.3702 & $1301.25\pm1.20$ & $-0.33\pm0.26$ & $1.26\pm1.07$ & $-1.05\pm0.19$ & [$-465\pm280$] \\ 
$m_B = 24.1$ & Si {\sc iv} & 1393.76018 & $1392.80\pm0.64$ & $-0.38\pm0.15$ & $1.27\pm0.56$ & $-1.19\pm0.19$ & $-206\pm138$ \\
		& Si {\sc iv} & 1402.77291 & $1401.92\pm0.62$ & $-0.50\pm0.74$ & $0.35\pm0.95$ & $-0.44\pm0.20$ & $-182\pm134$ \\
		& C {\sc iv} & 1548.204 & $1546.80\pm0.83$ & $-0.42\pm0.14$ & $2.00\pm0.85$ & $-2.09\pm0.19$ & [$-518\pm161$] \\
		& C {\sc iv} & 1550.781 & $1546.80\pm0.83$ & $-0.42\pm0.14$ & $2.00\pm0.85$ & $-2.09\pm0.19$ & [$-518\pm161$] \\
		& Al {\sc ii} & 1670.7886 & $1670.35\pm0.62$ & $-0.73\pm0.62$ & $0.39\pm0.36$ & $-0.71\pm0.20$ & $-96\pm112$ \\ \hline 
%%%%%%%%%%%%%%%%%
14212	& Si {\sc ii} & 1260.4221 & $1258.98\pm0.56$ & $-0.65\pm0.31$ & $0.93\pm0.48$ & $-1.52\pm0.24$ & $-343\pm133$ \\ 
EW(Ly$\alpha$)$=54.98$ [\AA] & O {\sc i} & 1302.1685 & $1301.56\pm0.74$ & $-0.52\pm0.28$ & $1.06\pm0.62$ & $-1.37\pm0.25$ & [$-394\pm172$] \\ 
$\Delta v_{\rm Ly\alpha}=188$ [km s$^{-1}$] & Si {\sc ii} & 1304.3702 & $1301.56\pm0.74$ & $-0.52\pm0.28$ & $1.06\pm0.62$ & $-1.37\pm0.25$ & [$-394\pm172$] \\ 
$m_B = 24.0$  & C {\sc ii} & 1334.5323 & $1333.42\pm0.55$ & $-0.69\pm0.16$ & $1.93\pm0.55$ & $-3.34\pm0.24$ & $-249\pm123$ \\
		& Si {\sc iv} & 1402.77291 & $1402.54\pm0.49$ & $-0.49\pm0.15$ & $1.30\pm0.43$ & $-1.59\pm0.24$ & $-50\pm105$ \\
		& Si {\sc ii} & 1526.70698 & $1525.49\pm0.58$ & $-0.86\pm0.38$ & $0.96\pm0.42$ & $-2.06\pm0.24$ & $-238\pm113$ \\
		& C {\sc iv} & 1548.204 & $1545.01\pm0.60$ & $-0.73\pm0.17$ & $2.32\pm0.66$ & $-4.27\pm0.30$ & [$-864\pm123$] \\
		& C {\sc iv} & 1550.781 & $1545.01\pm0.60$ & $-0.73\pm0.17$ & $2.32\pm0.66$ & $-4.27\pm0.30$ & [$-864\pm123$] \\
		& Fe {\sc ii} & 1608.45085 & $1606.52\pm1.00$ & $-0.32\pm0.23$ & $0.49\pm0.40$ & $-0.40\pm0.25$ & $-360\pm185$ \\ \hline
%%%%%%%%%%%%%%%%%
10600	& Si {\sc ii} & 1260.4221 & $1258.90\pm1.07$ & $-0.28\pm0.11$ & $2.34\pm1.18$ & $-1.65\pm0.19$ & $-363\pm254$ \\ 
EW(Ly$\alpha$)$=58.19$ [\AA] & O {\sc i} & 1302.1685 & $1298.58\pm0.39$ & $-0.61\pm0.47$ & $0.28\pm0.18$ & $-0.42\pm0.19$ & [$-1079\pm90$]\tablenotemark{a} \\ 
$\Delta v_{\rm Ly\alpha}=181$ [km s$^{-1}$] & Si {\sc ii} & 1304.3702 & $1298.58\pm0.39$ & $-0.61\pm0.47$ & $0.28\pm0.18$ & $-0.42\pm0.19$ & [$-1079\pm90$]\tablenotemark{a} \\ 
$m_B = 23.6$  & C {\sc ii} & 1334.5323 & $1333.38\pm0.45$ & $-0.54\pm0.28$ & $0.54\pm0.26$ & $-0.74\pm0.16$ & $-258\pm100$ \\
		& Si {\sc iv} & 1393.76018 & $1392.57\pm0.38$ & $-0.50\pm0.18$ & $0.74\pm0.26$ & $-0.92\pm0.16$ & $-256\pm82$ \\
		& Si {\sc iv} & 1402.77291 & $1401.64\pm0.46$ & $-0.74\pm0.35$ & $0.32\pm0.13$ & $-0.60\pm0.19$ & $-242\pm97$ \\
		& Si {\sc ii} & 1526.70698 & $1525.24\pm0.65$ & $-0.30\pm0.14$ & $1.00\pm0.50$ & $-0.75\pm0.18$ & $-289\pm127$ \\
		& C {\sc iv} & 1548.204 & $1546.75\pm0.36$ & $-0.92\pm0.32$ & $0.69\pm0.22$ & $-1.59\pm0.19$ & [$-528\pm71$]\tablenotemark{b} \\
		& C {\sc iv} & 1550.781 & $1546.75\pm0.36$ & $-0.92\pm0.32$ & $0.69\pm0.22$ & $-1.59\pm0.16$ & [$-528\pm71$]\tablenotemark{b} 
\enddata
\tablecomments{Columns: (1) Object ID. (2) Ion. (3) Wavelength in rest frame. (4) Observed wavelength of the line. (5) Amplitude of the emission line. (6) Width of the absorption line uncorrected for the instrumental broadening. (7) Equivalent width of the line. (8) Velocity offset of emission line relative to nebular emission lines.}
\tablenotetext{a}{The value of $\Delta v$ assumes that the rest wavelength of the blend is $1303.2694$ \AA.}
\tablenotetext{b}{The value of $\Delta v$ assumes that the rest wavelength of the blend is $1549.479$ \AA.}
\label{table_abs_line}
\end{deluxetable*}

\begin{figure}[t!]
  \begin{center}
    \includegraphics[width=80mm]{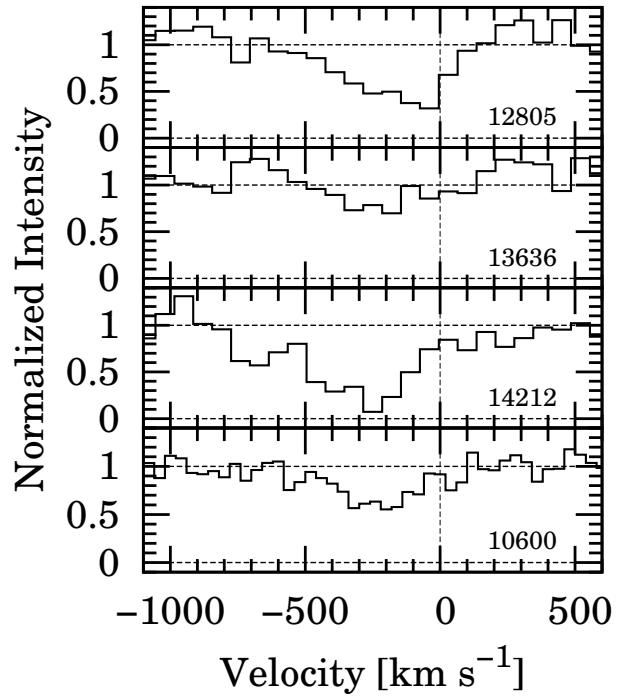}
  \end{center}
  \caption[IS absorption lines of LAEs]{{\footnotesize Average absorption line profiles of the four continuum-detected LAEs. These profiles are the average of the strong absorption lines, Si {\sc ii} $\lambda1260$, C {\sc ii} $\lambda1334$, and Si {\sc ii} $\lambda1526$ in the same manner as \citet{2013ApJ...779...52J}. The transitions of O {\sc i} $\lambda1302$ and Si {\sc ii} $\lambda1304$ are not used, since these lines are heavily blended owing to the low spectral resolution.}}
  \label{fig_average_absline}
\end{figure}

We derive the average absorption line profile of these strong transitions as a function of velocity. In the calculation, we do not use O {\sc i} $\lambda1302$ and Si {\sc ii} $\lambda1304$ transitions, since they could be heavily blended owing to the low spectral resolution. The derived average line profiles are shown in Figure \ref{fig_average_absline}. The covering fractions are estimated to be $\sim0.7$ for LAE 12805, $\sim0.3$ for 13636, $\sim0.9$ for 14212, and $\sim0.4$ for 10600 from the residual intensity in the core of the absorption line profiles. We additionally calculate the average depth of each best-fit Gaussian profile derived in Section \ref{subsec_measure_is_voffset}. This alternative is helpful to estimate adequately the depth of a profile with a low S/N. The values of $f_c$ are comparable to those derived from the average line profiles, with the exception of LAE 14212. The difference for LAE 14212 is because the $f_c$ of the average line profile is affected by a singular count of C {\sc ii} $\lambda1334$ line profile. 

The spectral resolution of our LRIS spectroscopy is $\sim4$ times lower than that of \citet{2013ApJ...779...52J}, preventing us from making a fair comparison between our LAEs and LBGs. We alternatively estimate EW of strong LIS absorption lines, EW(LIS), for our UV-continuum detected LAEs, and compare with results of composite LBG spectra in Section \ref{subsec_fc}.

\begin{deluxetable*}{cccccccc}
\setlength{\tabcolsep}{0.35cm} 
\tabletypesize{\scriptsize}
\tablecaption{Emission Line Features of UV-continuum Detected LAEs}
\tablehead{  \colhead{Object} & \colhead{Ion} & \colhead{$\lambda_{\rm rest}$} & \colhead{$\lambda_{\rm rest}^{\rm sys}$} & \colhead{$I/I_0$} & \colhead{$\sigma$} & \colhead{EW} &  \colhead{$\Delta v_{\rm IS}$} \\ 
\colhead{}& \colhead{}& \colhead{[\AA]}& \colhead{[\AA]}& \colhead{} &  \colhead{[\AA]} &  \colhead{[\AA]} &  \colhead{[km s$^{-1}$]}\\
\colhead{(1)}& \colhead{(2)}& \colhead{(3)}& \colhead{(4)} &  \colhead{(5)} &  \colhead{(6)} &  \colhead{(7)} & \colhead{(8)}}

\startdata
%\if0
12805	& Si {\sc ii}* & 1309.276 & $1309.26\pm0.44$ & $0.78\pm0.37$ & $0.82\pm0.45$ & $1.59\pm0.21$ & $5\pm100$ \\
		& O {\sc iii}$]$ & 1660.809 & $1660.58\pm0.51$ & $0.64\pm0.49$ & $0.41\pm0.31$ & $0.66\pm0.19$ & $-42\pm93$ \\
		& O {\sc iii}$]$ & 1666.150 & $1665.34\pm0.45$ & $0.89\pm0.76$ & $0.46\pm0.29$ & $1.02\pm0.19$ & $-147\pm81$ \\ \hline
%%%%%%%%%%%%%%%%%
13636	& Si {\sc ii}* & 1533.431 & $1532.97\pm0.62$ & $0.48\pm0.40$ & $0.44\pm0.39$ & $0.52\pm0.20$ & $-90\pm121$ \\
		& O {\sc iii}$]$ & 1666.150 & $1665.92\pm1.4$ & $0.39\pm0.30$ & $1.47\pm1.23$ & $1.45\pm0.19$ & $-42\pm257$ \\ \hline
%%%%%%%%%%%%%%%%%
14212	& Si {\sc ii}* & 1264.738 & $1265.06\pm0.49$ & $0.80\pm0.54$ & $0.65\pm0.45$ & $1.30\pm0.24$ & $76\pm117$ \\ \hline
%\fi
%%%%%%%%%%%%%%%%%
10600	& Si {\sc ii}* & 1264.738 & $1264.94\pm0.61$ & $0.45\pm0.32$ & $0.52\pm0.34$ & $0.59\pm0.16$ & $47\pm144$ \\
		& Si {\sc ii}* & 1533.431 & $1534.03\pm0.44$ & $0.73\pm0.48$ & $0.22\pm0.13$ & $0.41\pm0.18$ & $116\pm85$ \\
		& O {\sc iii}$]$ & 1666.150 & $1665.52\pm0.97$ & $0.37\pm0.27$ & $0.94\pm0.69$ & $0.86\pm0.20$ & $-113\pm175$ 
\enddata
\tablecomments{Columns: (1) Object ID. (2) Ion. (3) Wavelength in rest frame. (4) Observed wavelength of the line. (5) Amplitude of the emission line. (6) Width of the absorption line. (7) Equivalent width of the line. (8) Velocity offset of emission line relative to nebular emission lines.}
%\tablenotetext{a}{}
\label{table_emi_line}
\end{deluxetable*}

\begin{figure}[t!]
  \begin{center}    
    \includegraphics[width=85mm]{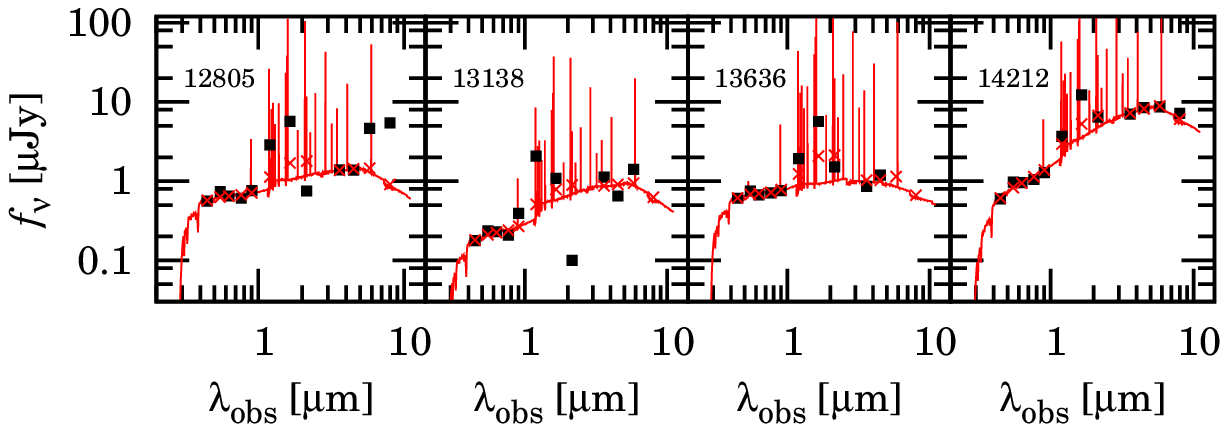}
    \includegraphics[width=85mm]{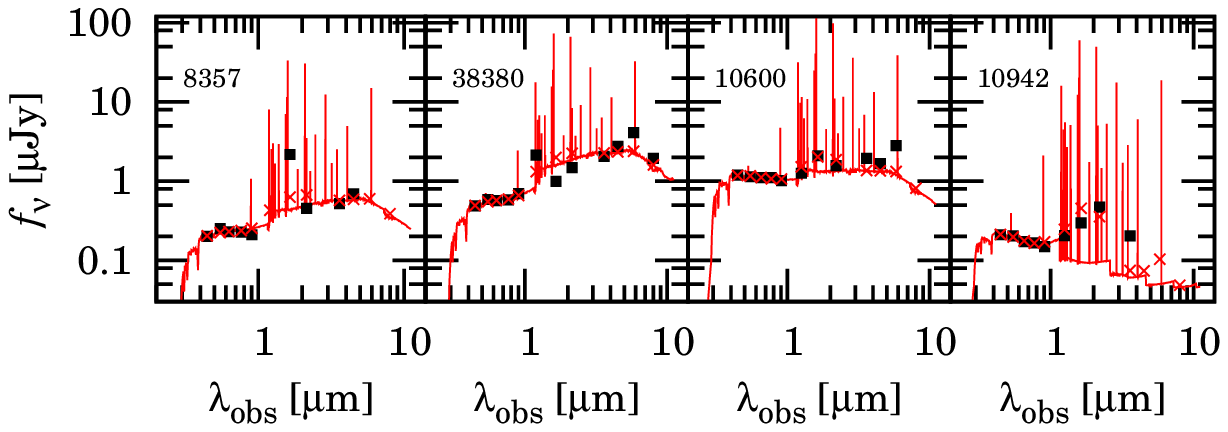}
  \end{center}
  \caption[HST images]{{\footnotesize Results of SED fitting for the eight LAEs with a $\Delta v_{\rm Ly\alpha}$ measurement. Red lines indicate the best-fit model spectra. Black filled squares represent observed magnitudes. Red crosses denote the flux densities at individual filters expected from the best-fit model spectra.}}
  \label{fig_bestfit_sed}
\end{figure}

%% SED FITTING
%%%%%%%%%%%%%%%%%%%%%%%%%%%%%%%%%%%%%%%%%%%%%%%%
\section{SED FITTING}\label{sec_sed_fitting}

In order to derive physical properties from stellar components, we perform SED fitting to the eight LAEs with known $z_{\rm sys}$. These LAEs have been imaged in several filters in the COSMOS or SXDS surveys. We use $B$, $V$, $r$, $i^\prime$, and $z^\prime$ data taken with Subaru/Suprime-Cam, $J$ data obtained with UKIRT/WFCAM, $K_s$ data from CFHT/WIRCAM \citep{2010ApJ...708..202M}, and {\it Spitzer}/IRAC 3.6, 4.5, 5.8, and 8.0 $\micron$ photometry from the Spitzer legacy survey of the UDS field.

The fitting procedure is the same as in \citet{2010MNRAS.402.1580O}. We create a spectral energy distribution (SED) of a starburst galaxy using a stellar population synthesis model, {\tt GALAXEV} \citep{2003MNRAS.344.1000B} including nebular emission \citep{2009A&A...502..423S}, with a Salpeter initial mass function with lower and upper mass cutoffs of $m_{\rm L} = 0.1 M_\odot$ and $m_{\rm U} = 100 M_\odot$. We assume a constant star formation history with a metallicity of $Z/Z_\odot = 0.2$. We use Calzetti's law \citep{2000ApJ...533..682C} for the stellar continuum extinction $E(B-V)$. These parameters are selected to be the same as those used in \citet{2013ApJ...765...70H} for consistency. The IGM absorption is applied to the spectra using the model of \citet{1995ApJ...441...18M}. The best-fit parameters and model spectra are shown in Table \ref{table_sed_fitting} and Figure \ref{fig_bestfit_sed}, respectively. The best-fit stellar mass of our LAEs ranges from $\log M_* \sim9$ to $\sim10$ which is broadly comparable to that of LBGs. This is because we choose bright objects from our LAE sample for the spectroscopic observations. Thus, the small $\Delta v_{\rm Ly\alpha}$ of LAEs does not appear to be caused by a difference in stellar mass between LAEs and LBGs.

%% RESULTS
%%%%%%%%%%%%%%%%%%%%%%%%%%%%%%%%%%%%%%%%%%%%%%%%
%%%%%%%%%%%%%%%%%%%%%%%%%%%%%%%%%%%%%%%%%%%%%%%%
\section{RESULTS}\label{sec_results}

%% VELOCITY OFFSET
%%%%%%%%%%%%%%%%%%%%%%%%%%%%%%%%%%%%%%%%%%%%%%%%
\subsection{Difference in $\Delta v_{\rm Ly\alpha}$ between LAEs and LBGs}\label{subsec_lya_voffset}

\begin{figure}[t]
  \begin{center}
    \includegraphics[width=80mm]{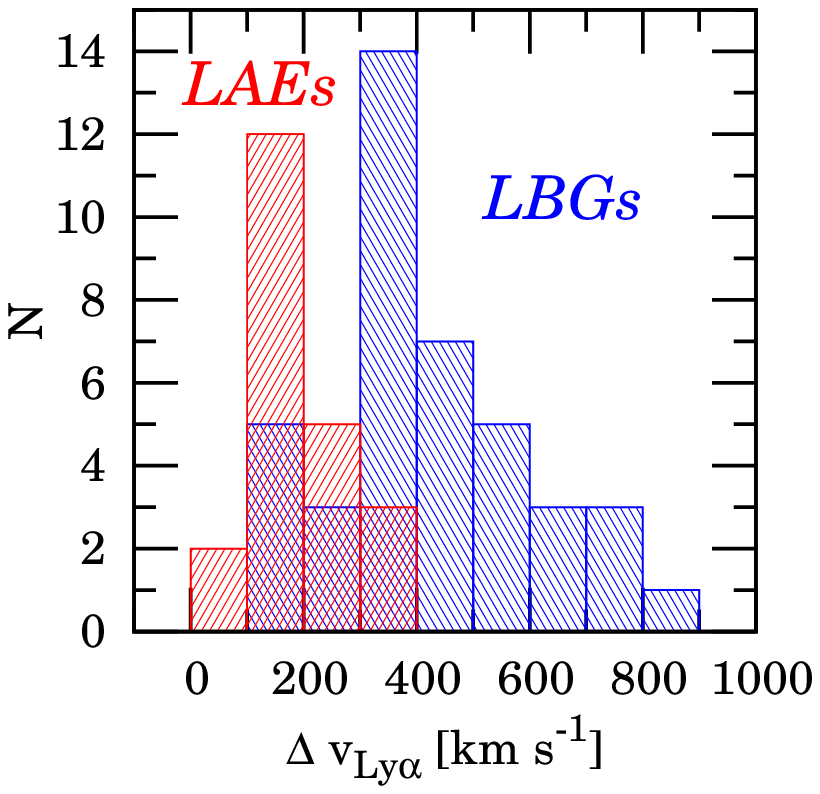}
  \end{center}
  \caption[IS absorption lines of LAEs]{{\footnotesize Histograms of Ly$\alpha$ velocity offset for the 22 LAEs in this study and literatures \citep{2011ApJ...730..136M,2011ApJ...729..140F,2013ApJ...765...70H,2013A&A...551A..93G,2013ApJ...775...99C}, and 41 LBGs given by \citet{2010ApJ...717..289S}.}}
  \label{fig_histogram_voffset_lya}
\end{figure}

In this section, we investigate statistically the difference in $\Delta v_{\rm Ly\alpha}$ between LAEs and LBGs in a compilation of LAEs with a $\Delta v_{\rm Ly\alpha}$ measurement in the previous studies including our 12 LAEs. The Ly$\alpha$ velocity offsets have previously been estimated for two objects in \citet{2011ApJ...730..136M}, three in the HETDEX survey \citep{2011ApJ...729..140F,2013ApJ...775...99C}, four from \citep{2013ApJ...765...70H}, and two LAEs in the MUSYC project \citep{2013A&A...551A..93G} at $z\sim2-3$. Among the objects in previous studies, COSMOS 13636 from \citep{2013ApJ...765...70H} is included in our sample of the 12 LAEs. We combine these 11 LAEs with our new 11, and construct a large sample consisting of 22 objects, which doubles the number of LAEs with $\Delta v_{\rm Ly\alpha}$ measurements. Figure \ref{fig_histogram_voffset_lya} shows the histogram of $\Delta v_{\rm Ly\alpha}$ using the newly-enlarged sample. This is the updated version of Figure 6 in \citet{2013ApJ...765...70H}. Similar to \citet{2013ApJ...765...70H}, we confirm that $\Delta v_{\rm Ly\alpha}$ of LAEs is systematically smaller than the values of LBGs. We carry out the non-parametric  Kolmogorv-Smirnov test for the two populations. The K-S probability is calculated to be $\sim10^{-7}$, indicating that $\Delta v_{\rm Ly\alpha}$ is definitively different between LBGs and LAEs. The weighted mean of the 22 objects is $\Delta v_{\rm Ly\alpha}= 234\pm 9$ km s$^{-1}$.

\begin{deluxetable}{cccccc}
\small
\tablewidth{0pt}
\tabletypesize{\scriptsize}
\tablecaption{SED Fitting Results for LAEs with a Systemic Redshift}
\tablehead{  \colhead{Slit Mask} & \colhead{Object} & \colhead{SFR} & \colhead{$E(B-V)$} & \colhead{$\log M_*$} & \colhead{$\chi^2_r$} \\ 
\colhead{}& \colhead{}& \colhead{[$M_\odot$ yr$^{-1}$]} &  \colhead{} & \colhead{[$M_\odot$]} & \\
\colhead{(1)}& \colhead{(2)}& \colhead{(3)}& \colhead{(4)} &  \colhead{(5)} &  \colhead{(6)}}

\startdata
COSMOS & 12805 & $34.7^{+1.3}_{-1.3}$ & $0.158^{+0.018}_{-0.018}$ & $9.442^{+0.134}_{-0.166}$ & $6.6$ \\ 
		& 13138 & $12.8^{+1.5}_{-1.6}$ & $0.185^{+0.035}_{-0.044}$ & $9.483^{+0.218}_{-0.197}$ & $1.7$ \\ 
		& 13636 & $67.9^{+1.2}_{-1.3}$ & $0.185^{+0}_{-0.009}$ & $9.051^{+0.115}_{-0.139}$ & $3.3$ \\ 
		& 14212 & $187.3^{+1.0}_{-1.1}$ & $0.326^{+0}_{-0.009}$ & $10.364^{+0.048}_{-0}$ & $13$ \\ 
		& 08357 & $9.52^{+2.1}_{-1.8}$ & $0.141^{+0.053}_{-0.053}$ & $9.213^{+0.277}_{-0.404}$ & $0.6$ \\ \hline
COSMOS3B & 38380 & $19.8^{+1.2}_{-1.1}$ & $0.132^{+0.018}_{-0.009}$ & $10.055^{+0.057}_{-0.111}$ & $1.4$ \\ \hline
SXDS495B & 10600 & $23.6^{+1.0}_{-1.1}$ & $0.053^{+0}_{-0.009}$ & $9.464^{+0.049}_{-0.041}$ & $4.7$ \\
		& 10942 & $14.9^{+33.5}_{-2.3}$ & $0.044^{+0.018}_{-0.018}$ & $7.734^{+0.110}_{-0.078}$ & $0.4$
%& 06713 & $4.93^{+1.1}_{-1.0}$ & $0^{+0.009}_{-0}$ & $9.450^{+0.092}_{-0.091}$ & $45$ \\ 
\enddata
\tablecomments{Columns: (1) Slit mask. (2) Object ID. (3) SFR. (4) Dust extinction. (5) Stellar mass. (6) Reduced $\chi^2$ of the SED fitting. }
\label{table_sed_fitting}
\end{deluxetable}

\begin{figure*}[t]
  \begin{center}
    \includegraphics[width=160mm]{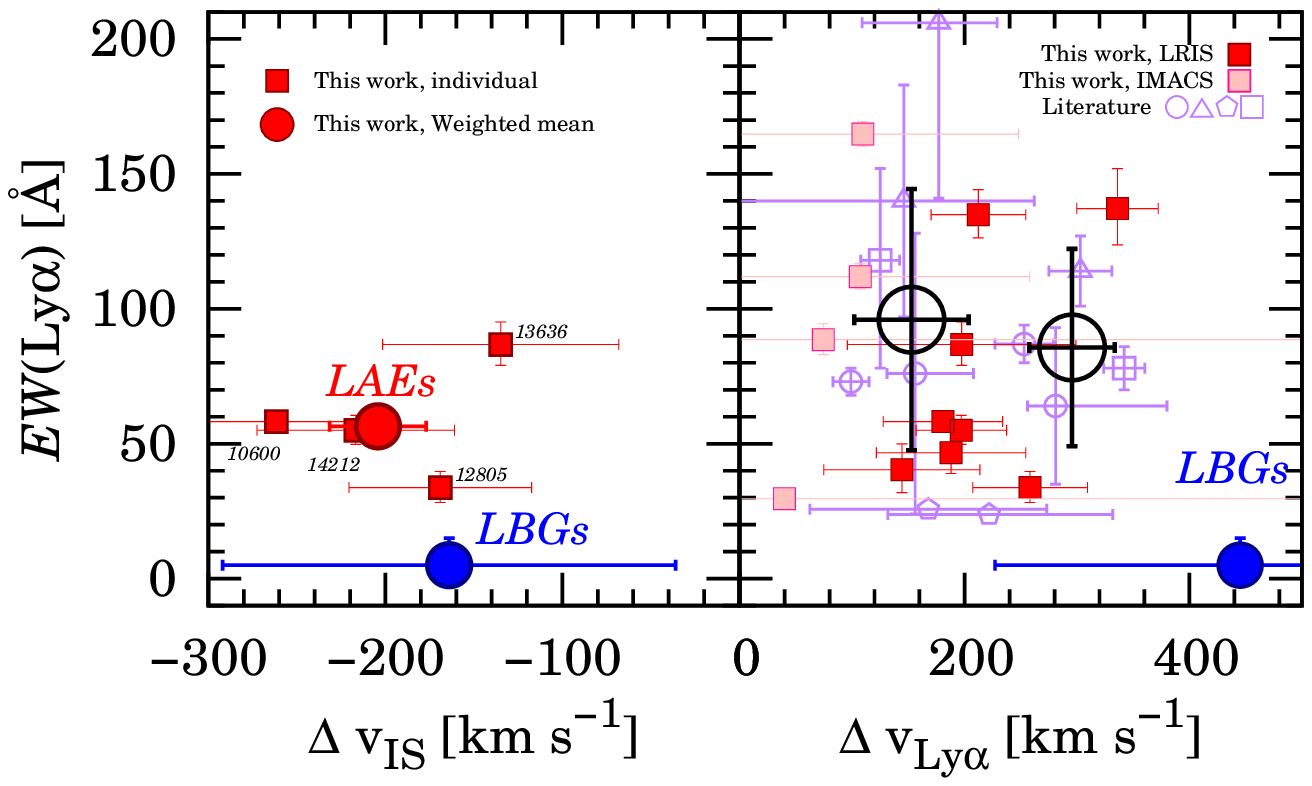}
  \end{center}
  \caption[velocity offset ]{{\footnotesize Rest-frame Ly$\alpha$ EW as a function of $\Delta v_{\rm Ly\alpha}$ (right) and $\Delta v_{\rm IS}$ (left). The red and magenta squares indicates LAEs observed with Keck/LRIS and Magellan/IMACS, respectively. The purple open symbols denote LAEs in the previous studies, \citep[squares; ][]{2011ApJ...730..136M}, \citep[triangles; ][]{2011ApJ...729..140F,2013ApJ...775...99C}, \citep[circles; ][]{2013ApJ...765...70H}, and \citep[pentagons; ][]{2013A&A...551A..93G}. The black open circles with error bars represent the average EW in each $\Delta v_{\rm Ly\alpha}$ bin. The large red circle in the left panel depicts the weighted mean of the four continuum-detected LAEs. The blue symbol denotes the average of 41 LBGs, with the error bars corresponding to the $68$ percentiles of the $\Delta v_{\rm Ly\alpha}$ and $\Delta v_{\rm IS}$ distributions \citep{2010ApJ...717..289S} and the EW distribution \citep{2008ApJS..175...48R}. }}
  \label{fig_voffset_lya_ism_ew}
\end{figure*}

We plot EW(Ly$\alpha$) and $\Delta v_{\rm Ly\alpha}$ of our new sample in the right panel of Figure \ref{fig_voffset_lya_ism_ew}. We confirm the anti-correlation between EW(Ly$\alpha$) and $\Delta v_{\rm Ly\alpha}$ in high-$z$ star-forming galaxies including objects with a high Ly$\alpha$ EW ({\it Hashimoto relation}) suggested by \citet{2013ApJ...765...70H}. The larger sample clarifies that $\Delta v_{\rm Ly\alpha}$ decreases with increasing EW(Ly$\alpha$). The similar trend has been found for UV-continuum selected galaxies at $z\sim2-3$. \citet{2003ApJ...588...65S} have calculated velocity offsets between Ly$\alpha$ emission and IS absorption ($\Delta v_{\rm em-abs}$) for composite spectra of LBGs, and have investigated relation between Ly$\alpha$ EW and $\Delta v_{\rm em-abs}$. Four LBG subsamples divided according to their Ly$\alpha$ EW reveal a trend that Ly$\alpha$ EW increases with decreasing $\Delta v_{\rm em-abs}$ in the EW range of $-15 - +50$\,\AA. The $\Delta v_{\rm em-abs}$ of our UV-continuum detected LAEs is very consistent with the trend of \citet{2003ApJ...588...65S}, as shown in the top panel of Figure \ref{fig_ewlya_dvemabs_ewlis}. Nevertheless, Ly$\alpha$ and IS velocity offsets from $z_{\rm sys}$ would be capable of distinguishing effects of Ly$\alpha$ resonant scattering and galactic outflow on $\Delta v_{\rm em-abs}$.

\begin{deluxetable*}{cccccccc}
\setlength{\tabcolsep}{0.5cm} 
\tabletypesize{\scriptsize}
\tablecaption{Properties of the NB-selected Galaxies with detections of Ly$\alpha$ and Nebular Emission lines in the Previous Studies}
\tablehead{\colhead{Object} & \colhead{$z_{\rm sys}$} & \colhead{EW(Ly$\alpha$)} & \colhead{$\Delta v_{\rm Ly\alpha}$} & \colhead{SFR} & \colhead{$E(B-V)$} & \colhead{$\log M_*$} & \colhead{Comments} \\ 
\colhead{}& \colhead{}& \colhead{[\AA]} & \colhead{[km s$^{-1}$]} & \colhead{[$M_\odot$ yr$^{-1}$]} & \colhead{} & \colhead{[$M_\odot$]} & \colhead{} \\
\colhead{(1)}& \colhead{(2)}& \colhead{(3)}& \colhead{(4)} &  \colhead{(5)} & \colhead{(6)}& \colhead{(7)} &  \colhead{(8)}}

\startdata
\multicolumn{8}{c}{\citet{2011ApJ...730..136M}} \\ \hline
LAE27878 & 3.11879 & $118^{+34}_{-40}$ & $125\pm17.3$ & $32^{+9}_{-7}$ & \nodata & $9.97^{+0.378}_{-0.395}$\tablenotemark{a} & $[$O {\sc iii}$]\lambda5007$ \\ 
LAE40844 & 3.11170 & $78^{+8}_{-8}$ & $342\pm18.3$ & $113^{+120}_{-60}$ & \nodata & $9.80^{+0.734}_{-0.363}$\tablenotemark{a} & \\ \hline
\multicolumn{8}{c}{\citet{2011ApJ...729..140F} and \citet{2013ApJ...775...99C}} \\ \hline
HPS 194 & 2.28628 & $114\pm13$ & $303\pm28$ & $>29.3$\tablenotemark{b} & $0.09\pm0.06$ & $10.2^{+0.08}_{-0.14}$ & HETDEX sample \\ 
HPS 256 & 2.49024 & $206\pm65$ & $177^{+52}_{-68}$ & $>35.4$\tablenotemark{b} & $0.10\pm0.09$ & $8.28^{+0}_{-0.02}$ & H$\beta$, $[$O {\sc iii}$]\lambda4959$, $[$O {\sc iii}$]\lambda5007$, H$\alpha$ \\ 
HPS 251 & 2.28490 & $140\pm43$ & $146^{+116}_{-156}$ & $>9.9$\tablenotemark{b} & $0.07\pm0.08$ & $9.04^{+0.73}_{-0.04}$ & \\ \hline
\multicolumn{8}{c}{\citet{2013ApJ...765...70H}} \\ \hline
CDFS-3865 & 2.17210 & $64^{+29}_{-29}$ & $281^{+99}_{-25}$ & $190^{+13}_{-13}$\tablenotemark{b} & $0.185^{+0.009}_{-0.009}$ & $9.50^{+0.028}_{-0.018}$ & Subaru NB387 sample \\ 
CDFS-6482 & 2.20443 & $76^{+52}_{-52}$ & $156^{+52}_{-25}$ & $48^{+10}_{-9}$\tablenotemark{b} & $0.185^{+0.026}_{-0.018}$ & $9.72^{+0.087}_{-0.071}$ & $[$O {\sc iii}$]\lambda5007$, H$\alpha$ \\ 
COSMOS-13636 & 2.16125\tablenotemark{c} & $73^{+5}_{-5}$ & $99^{+16}_{-16}$\tablenotemark{c} & $18^{+3}_{-3}$\tablenotemark{b} & $0.273^{+0.018}_{-0.079}$ & $9.30^{+0.078}_{-0.330}$ & \\ 
COSMOS-30679 & 2.19776 & $87^{+7}_{-7}$ & $253^{+26}_{-26}$ & $45^{+5}_{-5}$\tablenotemark{b} & $0.528^{+0.026}_{-0.026}$ & $10.3^{+0.124}_{-0.151}$ & \\ \hline
\multicolumn{8}{c}{\citet{2013A&A...551A..93G}} \\ \hline
LAE27 & 3.0830 & $25.7$ & $167.8\pm105.3$ & \nodata & $\lesssim 0.1$ & $9.95^{+0.13}_{-0.17}$ & MUSYC sample \\ 
z3LAE2 & 3.1118 & $23.8$ & $221.8\pm90.0$ & \nodata & $0.32^{+0.06}_{-0.23}$ & $9.95^{+0.13}_{-0.17}$ & H$\beta$, $[$O {\sc iii}$]\lambda4959$, $[$O {\sc iii}$]\lambda5007$
\enddata
\tablecomments{Columns: (1) Object ID. (2) Systemic redshift. (3) Ly$\alpha$ equivalent width. (4) Ly$\alpha$ velocity offset. (5) SFR. (6) Dust extinction. (7) Stellar mass. (8) Comments. }
\tablenotetext{a}{Estimated in \citet{2014ApJ...780...20R}.}
\tablenotetext{b}{Based on H$\alpha$ flux.}
\tablenotetext{c}{The $\Delta v_{\rm Ly\alpha}$ of this object is calculated to be $197\pm102$ km s$^{-1}$ in our FMOS observation with higher spectral resolution than that of the Keck-II/NIRSPEC spectrosocpy in \citet{2013ApJ...765...70H} (see Table \ref{table_keck_lae}). }
\label{table_previous_works}
\end{deluxetable*}

%% OUTFLOW VELOCITY
%%%%%%%%%%%%%%%%%%%%%%%%%%%%%%%%%%%%%%%%%%%%%%%%
\subsection{Difference in $\Delta v_{\rm IS}$ between LAEs and LBGs}\label{subsec_is_voffset}

We additionally examine a possible difference in $\Delta v_{\rm IS}$ between LAEs and LBGs. The weighted means of $\Delta v_{\rm IS}$ of the absorption lines are calculated to be $-134\pm67$, $-261\pm48$, $-216\pm56$, and $-169\pm52$ km s$^{-1}$ for LAE 13636, 10600, 14212, and 12805, respectively. In the calculation of average $\Delta v_{\rm IS}$ for each object, we exclude several line-pairs with a large $\Delta v_{\rm IS}$ of $\lesssim-500$ km s$^{-1}$ which are not reliably determined due to a line blending. As shown in Table \ref{table_abs_line}, we find that almost all IS absorption lines are blueshifted with respect to $z_{\rm sys}$ by $\lesssim -200$ km s$^{-1}$, which indicates that gaseous outflows are present in the continuum-detected LAEs. 

The left panel in Fig. \ref{fig_voffset_lya_ism_ew} represents the relation between EW(Ly$\alpha$) and $\Delta v_{\rm IS}$. The average of the four is $\Delta v_{\rm IS}=-204\pm27$ km s$^{-1}$, which is comparable to that of LBGs \cite[e.g., ][]{2006ApJ...646..107E,2010ApJ...717..289S} in contrast to $\Delta v_{\rm Ly\alpha}$, although the current small sample of LAEs with a $\Delta v_{\rm IS}$ is insufficient to provide a definitive conclusion on $\Delta v_{\rm IS}$ of LAEs and LBGs.

%% HI Covering Fraction 
%%%%%%%%%%%%%%%%%%%%%%%%%%%%%%%%%%%%%%%%%%%%%%%%
\subsection{Difference in $f_c$ between LAEs and LBGs}\label{subsec_fc}

\begin{figure}[t!]
  \begin{center}
    \includegraphics[width=80mm]{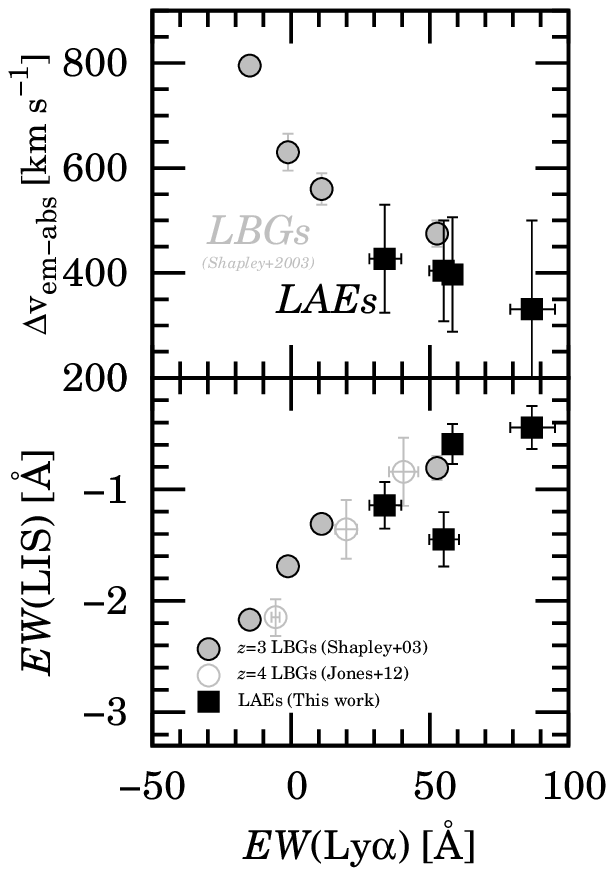}
  \end{center}
  \caption[IS absorption lines of LAEs]{{\footnotesize Top: velocity offset between Ly$\alpha$ emission and LIS absorption lines as a function of EW(Ly$\alpha$). Bottom: equivalent width of LIS absorption lines as a function of EW(Ly$\alpha$). The black filled squares indicate the UV-continuum detected LAEs. The gray open and filled circles denote composite LBG spectra at $z\sim3$ \citep{2003ApJ...588...65S} and $z\sim4$ \citep{2012ApJ...751...51J}, respectively. The EW(LIS) of LAEs is the average equivalent width of six strong LIS absorption lines, Si {\sc ii} 1260, O {\sc i}+Si {\sc ii} 1303, C {\sc ii} 1334, Si {\sc ii} 1526, Fe {\sc ii} 1608, and Al {\sc ii} 1670. }}
  \label{fig_ewlya_dvemabs_ewlis}
\end{figure}

\begin{figure}[t]
  \begin{center}
    \includegraphics[width=80mm]{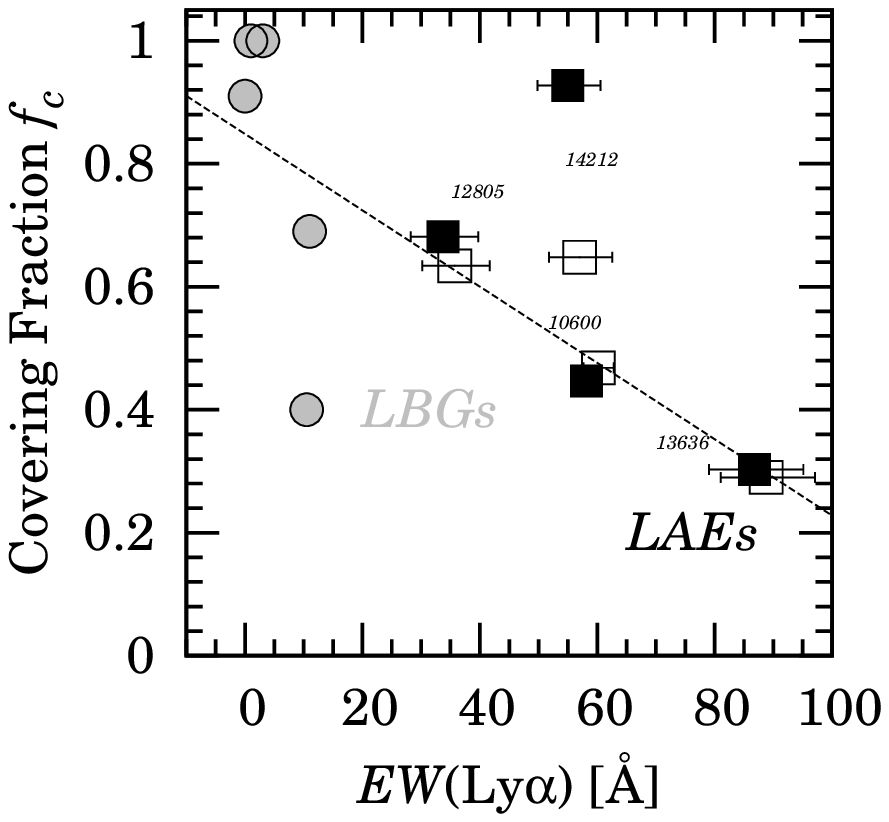}
  \end{center}
  \caption[IS absorption lines of LAEs]{{\footnotesize Covering fraction of H {\sc i} gas, $f_c$, as a function of Ly$\alpha$ EW. The $f_c$ values of the four LAEs are estimated from the depth of the average values in the individual profile-fitting (open squares; Table \ref{table_abs_line}) and average LIS absorption line profiles (filled squares; Fig. \ref{fig_average_absline}). The gray circles indicate LBGs at $z=2-3$ in \citet{2013ApJ...779...52J}. The dashed line denotes a linear fit to the data.}}
  \label{fig_fc_ew}
\end{figure}

We compare the H {\sc i} covering fraction $f_c$ of LAEs derived in Section \ref{subsec_measure_fc} with that of $z\sim2-3$ LBGs in \citet{2013ApJ...779...52J}. Note that we here place lower limits on $f_c$ when $\tau\gg1$. Figure \ref{fig_fc_ew} displays the relation between $f_c$ and Ly$\alpha$ EW, indicating a tentative trend that $f_c$ decreases with Ly$\alpha$ EW. This trend has already been found in \citet{2013ApJ...779...52J} using an LBG sample. We find that the trend continues into objects with a higher Ly$\alpha$ EW. Our slope of the trend is slightly steeper than that in \citet{2013ApJ...779...52J}, which would result from the wider dynamic range in Ly$\alpha$ EW. However, this trend could arise from the difference in the spectral resolution, although this tendency may marginally be found for LAEs alone. In addition to $f_c$, we compare EW(LIS) between LAEs and $z=3-4$ LBGs in the bottom panel of Figure \ref{fig_ewlya_dvemabs_ewlis}. \citet{2003ApJ...588...65S} have found that EW(LIS) decreases with increasing EW(Ly$\alpha$) with composite spectra of LBGs. Our LAEs with EW(Ly$\alpha)=30-90$\,\AA\, follow the trend between EW(Ly$\alpha$) and EW(LIS), which might be indicative of a low velocity dispersion and/or low $f_c$, as suggested by \citet{2003ApJ...588...65S}. These results related to the low $f_c$ imply the need for modeling Ly$\alpha$ line profiles emitted from a non-spherical shell of neutral gas \citep[e.g., ][]{2013arXiv1308.1405Z,2014A&A...563A..77B}.

%% DISCUSSION
%%%%%%%%%%%%%%%%%%%%%%%%%%%%%%%%%%%%%%%%%%%%%%%%
%%%%%%%%%%%%%%%%%%%%%%%%%%%%%%%%%%%%%%%%%%%%%%%%
%%%%%%%%%%%%%%%%%%%%%%%%%%%%%%%%%%%%%%%%%%%%%%%%
\section{DISCUSSION} \label{sec_discussion}

%% ORIGIN OF LYA VELOCITY OFFSET
%%%%%%%%%%%%%%%%%%%%%%%%%%%%%%%%%%%%%%%%%%%%%%%%
\subsection{Origin of Small $\Delta v_{\rm Ly\alpha}$ in LAEs}\label{subsec_origin}

As described in the previous sections, we definitely confirm the anti-correlation between $\Delta v_{\rm Ly\alpha}$ and Ly$\alpha$ EW by using a larger LAE sample than previously available. In this section, we explore the physical origin of the small $\Delta v_{\rm Ly\alpha}$ in high Ly$\alpha$ EW galaxies. 

Models predict that the redshift of the Lya emission line should increase with either outflow velocity or neutral hydrogen column density ($N_{\rm HI}$) \citep{2006A&A...460..397V, 2008A&A...491...89V}. We have shown that the outflow velocities of LAE are comparable to those of LBGs, so the smaller $\Delta v_{\rm Ly\alpha}$ for LAE is likely to be due to lower column densities in these objects. 

In order to address the origin of the small $\Delta v_{\rm Ly\alpha}$ in LAEs, we introduce the {\it velocity offset ratio}, 

\begin{equation}
  R^{\rm Ly\alpha}_{\rm IS} \equiv \frac{\Delta v_{\rm Ly\alpha}}{\Delta v_{\rm IS}}. 
\end{equation}

\noindent The value of $R^{\rm Ly\alpha}_{\rm IS}$ could trace purely physical properties such as $N_{\rm HI}$ and the dust amount by excluding the kinematic effect of a bulk outflow, since the quantity is normalized by the outflowing velocity, as suggested in \citep{2006A&A...460..397V}. \citet{2013ApJ...765...70H} infer the average value of $\Delta v_{\rm IS}$ for LAEs from a stacked spectrum of four LAEs with a $z_{\rm sys}$, and compare $R^{\rm Ly\alpha}_{\rm IS}$ between LAEs and LBGs. In the stacking analysis, $R^{\rm Ly\alpha}_{\rm IS}$ is found to be $\sim1$ for LAEs which is slightly-smaller than that of LBGs, but the uncertainties are large. 

Here, we estimate $R^{\rm Ly\alpha}_{\rm IS}$ for the four continuum-detected LAEs, and compare the quantities with those of $z=2-3$ LBGs in \citet{2006ApJ...647..128E,2006ApJ...646..107E}. In the comparison, we use LBGs with a negative $\Delta v_{\rm IS}$ value that indicates the outflow is present. The LAEs have a $R^{\rm Ly\alpha}_{\rm IS}$ of $\sim0.6-1.4$, while LBGs have a wide variety of the quantity from $0$ to $\sim10$. Nevertheless, the average $R^{\rm Ly\alpha}_{\rm IS}$ for the LAEs is systematically smaller than that of LBGs. 
This indicates that LAEs tend to have a small $N_{\rm HI}$ compared to LBGs based on the expanding gas shell model of \citet{2006A&A...460..397V}. The small $R^{\rm Ly\alpha}_{\rm IS}$ in LAEs would be indicative of a small $N_{\rm HI}$ in LAEs. 

\begin{figure*}[t]
  \begin{center}
    \includegraphics[width=130mm]{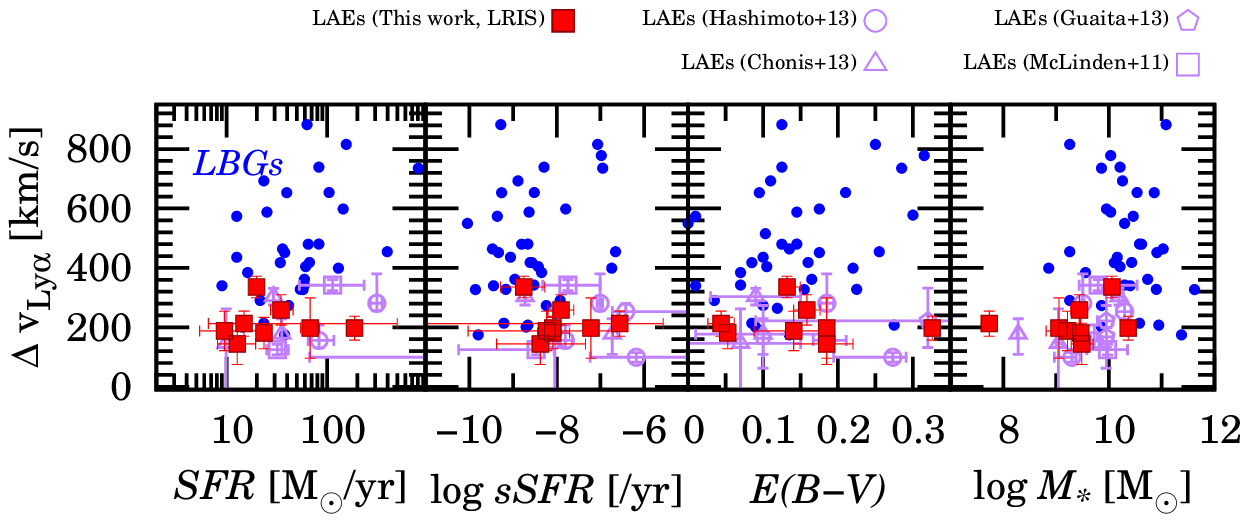}
  \end{center}
  \caption[IS absorption lines of LAEs]{{\footnotesize Correlations between $\Delta v_{\rm Ly\alpha}$ and physical properties inferred from the SED fitting. The symbols are the same as Figure \ref{fig_voffset_lya_ism_ew}. We multiply the physical quantities of LBGs in \citet{2006ApJ...647..128E} by 1.8, because they use a Chabrier IMF \citep{2003PASP..115..763C} in the SED fitting.}}
  \label{fig_correlation_dvlya}
\end{figure*}

\begin{figure*}[t]
  \begin{center}
    \includegraphics[width=130mm]{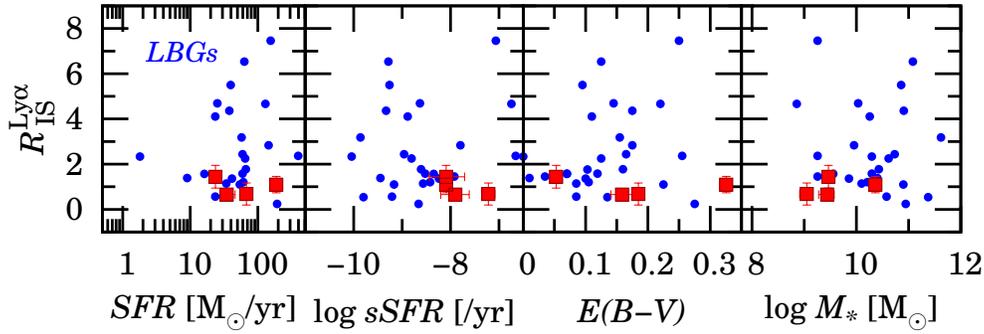}
  \end{center}
  \caption[IS absorption lines of LAEs]{{\footnotesize Same as Figure \ref{fig_correlation_dvlya}, but for $R_{\rm IS}^{\rm Ly\alpha}$. Red squares indicate the four UV continuum-detected LAEs. }}
  \label{fig_correlation_vratio}
\end{figure*}

Next, we examine possible correlations of $\Delta v_{\rm Ly\alpha}$ and $R_{\rm IS}^{\rm Ly\alpha}$ with physical properties inferred from the SED fitting (Section \ref{sec_sed_fitting}). In correlation tests, $\Delta v_{\rm Ly\alpha}$ and $R_{\rm IS}^{\rm Ly\alpha}$ correlate most strongly with mass-related quantities, and SFR, respectively. Figures \ref{fig_correlation_dvlya} and \ref{fig_correlation_vratio} show the correlations of these quantities, respectively, including LAEs and LBGs with a $z_{\rm sys}$ in the literatures. The SFR value of several LAEs is based on a H$\alpha$ flux through the relation of \citet{1998ARA&A..36..189K}. The SFR based on a H$\alpha$ flux is found to be comparable to the value inferred from SED fitting \citep{2013ApJ...765...70H}. We conduct Spearman rank correlation tests in order to find the most related physical quantities to $\Delta v_{\rm Ly\alpha}$ and $R_{\rm IS}^{\rm Ly\alpha}$ in the same manner as \citet{2010ApJ...717..289S}. Table \ref{table_correlation} summarizes the results of the Spearman rank correlation tests.

\begin{deluxetable}{crcrc}
\setlength{\tabcolsep}{0.5cm} 
\tabletypesize{\scriptsize}
\tablecaption{Correlations between Ly$\alpha$ Kinematics and Galaxy Properties}
\tablehead{\colhead{Quantity} & \colhead{$\Delta v_{\rm Ly\alpha}$} & \colhead{$N_{\Delta v_{\rm Ly\alpha}}$} & \colhead{$R^{\rm Ly\alpha}_{\rm IS}$} & \colhead{$N_{R^{\rm Ly\alpha}_{\rm IS}}$}\\ 
\colhead{(1)}& \colhead{(2)}& \colhead{(3)}& \colhead{(4)}& \colhead{(5)}}

\startdata
SFR & $0.065$ & $53$ & $0.241$ & $31$ \\ %, 0.8858 (33)$ \\ 
sSFR & $-0.098$ & $51$ & $-0.852$ & $31$ \\ %, -0.3798 (33)$ \\
$E(B-V)$ & $0.792$ & $54$ & $0.272$ & $34$ \\ %, 0.5617 (36)$ \\ 
$M_*$ & $0.001$ & $52$ & $0.810$ & $31$ %, 0.4471 (33)$ 
\enddata

\tablecomments{Columns: (1) Physical quantity. (2) Probabilities satisfying the null hypothesis that the quantities are not correlated in Spearman rank correlation tests. A smaller absolute value of the probabilities implies that a physical property more correlates with a Ly$\alpha$ velocity offset. Negative values indicates anti-correlations. (3) Number of galaxies in the correlation test between $\Delta v_{\rm Ly\alpha}$ and physical quantities. (4)-(5) Probabilities and galaxy numbers in the correlation tests for $R^{\rm Ly\alpha}_{\rm IS}$. Two LBGs with an extremely high $R^{\rm Ly\alpha}_{\rm IS}$ value of $>25$ are excluded in the correlation tests.}
\label{table_correlation}
\end{deluxetable}

\begin{figure*}[t!]
  \begin{center}
    \includegraphics[width=160mm]{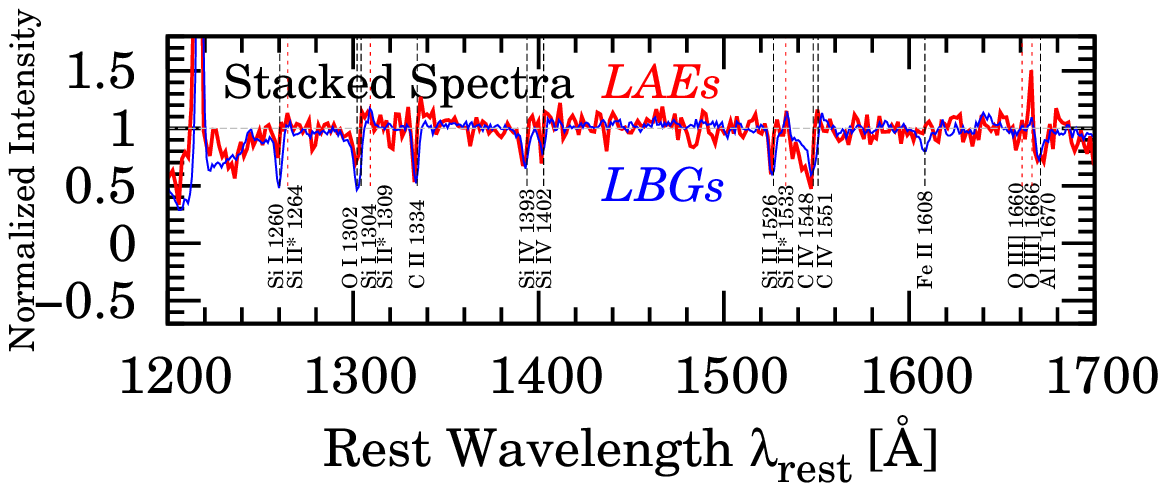}
  \end{center}
  \caption[IS absorption lines of LAEs]{{\footnotesize Composite rest-frame UV spectra of the eight LAEs observed with Keck/LRIS (red) and LBGs in \citet{2003ApJ...588...65S} (blue). The spectra are normalized to unity in the continuum levels. Black and red vertical dashed lines indicates wavelengths of IS absorption and emission lines, respectively.}}
  \label{fig_stack}
\end{figure*}

For Ly$\alpha$ velocity offsets, we find that the $\Delta v_{\rm Ly\alpha}$ strongly correlates with SFR and stellar mass, which has not been observed previously in an LBG sample \citep{2010ApJ...717..289S}. These correlations may have merged because our sample covers larger dynamic ranges of SFR and $M_*$. The correlation between $\Delta v_{\rm Ly\alpha}$ and sSFR may arise from the stellar mass. 

As far as the velocity offset ratio is concerned, we do not find a notable correlation between $R_{\rm IS}^{\rm Ly\alpha}$ and the physical properties. Nonetheless, the correlation tests indicate that $R_{\rm IS}^{\rm Ly\alpha}$ most correlates with SFR among the four physical quantities. The correlation may reflect the connection between star formation and $N_{\rm HI}$, if $R_{\rm IS}^{\rm Ly\alpha}$ is sensitive to $N_{\rm HI}$. A larger sample of LAEs with a $R_{\rm IS}^{\rm Ly\alpha}$ measurement might reveal its physical connections with galactic properties.

%% ORIGIN
%%%%%%%%%%%%%%%%%%%%%%%%%%%%%%%%%%%%%%%%%%%%%%%%
%%%%%%%%%%%%%%%%%%%%%%%%%%%%%%%%%%%%%%%%%%%%%%%%
\subsection{What is the Physical Origin of Strong Ly$\alpha$ Emission?}\label{subsec_discuss_origin}

With our larger sample of LAEs with a $z_{\rm sys}$, we confirm conclusively that LAEs typically have a smaller $\Delta v_{\rm Ly\alpha}$ than LBGs with a lower Ly$\alpha$ EW, while their outflowing velocities are similar in the two populations. These results yield a small $R^{\rm Ly\alpha}_{\rm IS}$ in LAEs, which indicates a small $N_{\rm HI}$ in galaxies with a high Ly$\alpha$ EW. The anti-correlations of $f_c$ and EW(LIS) with Ly$\alpha$ EW in Figures \ref{fig_fc_ew} and \ref{fig_ewlya_dvemabs_ewlis} are consistent with the small $N_{\rm HI}$ in LAEs. The patchy H {\sc i} gas clouds surrounding the central source would lead to a small flux-averaged $N_{\rm HI}$ corresponding to a small $R^{\rm Ly\alpha}_{\rm IS}$. In this condition, Ly$\alpha$ photons could easily escape less affected by resonant scattering in the clouds. The results of our kinematic analyses support the idea that the H {\sc i} column density is a key quantity determining Ly$\alpha$ emissivity. 

Moreover, recent NIR spectroscopy by \citet{2013ApJ...769....3N} has suggested that LAEs have a large $[$O {\sc iii}$]/[$O {\sc ii}$]$ ratio, indicating these systems are highly ionized with density-bounded H {\sc ii} regions. This tendency has been confirmed by a subsequent systematic study in \citet{2013arXiv1309.0207N}. The large $[$O {\sc iii}$]/[$O {\sc ii}$]$ ratio also indicates a low column density of H {\sc i} gas. A stacked UV continuum spectrum of our eight LAEs shows that LIS absorption lines have a low EW, as shown in Figures \ref{fig_stack} (see also Fig. \ref{fig_ewlya_dvemabs_ewlis}). The weak LIS absorption lines are consistent with a large $[$O {\sc iii}$]/[$O {\sc ii}$]$ ratio in LAEs \citep[e.g., ][]{2012ApJ...751...51J}. 

In our first paper of the series investigating LAE structures, we find that LAEs with a high Ly$\alpha$ EW tend to be a non-merger, to show a small Ly$\alpha$ spatial offset between Ly$\alpha$ and stellar continuum emission $\delta_{\rm Ly\alpha}$, and to have a small ellipticity by using a large sample of 426 LAEs \citep{2014ApJ...785...64S}. On the basis of these results on the gas distribution, the difference in H {\sc i} column density explains the Ly$\alpha$-EW dependences of the merger fraction, the Ly$\alpha$ spatial offset, and the galaxy inclination. For objects with density-bounded H {\sc ii} regions, Ly$\alpha$ photons would directly escape from central ionizing sources, which produce a small $\delta_{\rm Ly\alpha}$. The low H {\sc i} abundance along the line of sight also induces the preferential escape of Ly$\alpha$ to the face-on direction. 

All of the above results suggest that ionized regions with small amounts of H {\sc i} gas dominate in galaxies with a high Ly$\alpha$.

%% CONCLUSION
%%%%%%%%%%%%%%%%%%%%%%%%%%%%%%%%%%%%%%%%%%%%%%%%
%%%%%%%%%%%%%%%%%%%%%%%%%%%%%%%%%%%%%%%%%%%%%%%%
\section{SUMMARY and CONCLUSION}\label{sec_conclusion}

We carry out deep optical spectroscopy for our large sample of LAEs at $z=2.2$ in order to detect their Ly$\alpha$ lines with Keck/LRIS. We compare redshifts of the Ly$\alpha$ and nebular emission lines detected with Subaru/FMOS, and calculate $\Delta v_{\rm Ly\alpha}$ for new 11 LAEs. This observation doubles the sample size of LAEs with a $\Delta v_{\rm Ly\alpha}$ measurement in literatures. 
 
The conclusions of this study are summarized below. 
 
\begin{itemize}
  \item Almost all of our new LAEs have a $\Delta v_{\rm Ly\alpha}$ of $\sim200$ km s$^{-1}$ which is systematically-smaller than that of LBGs. Using 22 LAEs with $\Delta v_{\rm Ly\alpha}$ measurements taken from our new observations and the literature, we definitively confirm the anti-correlation between Ly$\alpha$ EW and $\Delta v_{\rm Ly\alpha}$ suggested by previous work. 
  \item Long exposure times and the high sensitivity of LRIS at blue wavelengths enabled us to successfully detect IS absorption lines against faint UV continua from four individual LAEs. These IS absorption lines are found to be blueshifted from the systemic redshift by $200-300$ km s$^{-1}$, indicating strong gaseous outflows are present even in LAEs. 
  \item We estimate $R_{\rm IS}^{\rm Ly\alpha}$ ($\equiv\Delta v_{\rm Ly\alpha}/\Delta v_{\rm IS}$) that would be a quantity sensitive to $N_{\rm HI}$ for the four UV continuum-detected LAEs. We find the value of $R_{\rm IS}^{\rm Ly\alpha}$ in LAEs to be smaller than that of LBGs, indicating a lower $N_{\rm HI}$ in LAEs. We performed a test for correlations between $R_{\rm IS}^{\rm Ly\alpha}$ and physical properties inferred from SED fitting. As a result, we tentatively conclude that SFR may be most closely related to $R_{\rm IS}^{\rm Ly\alpha}$. The correlation may suggest that the star formation preferentially occurs in systems with large amounts of neutral hydrogen gas, which would have a larger value of $R_{\rm IS}^{\rm Ly\alpha}$. 
  \item We estimate the covering fraction, $f_c$, of surrounding H {\sc i} gas from the depth of LIS absorption lines the four LAEs. We identify a tentative trend for $f_c$ to decrease with increasing Ly$\alpha$ EW, as suggested by a study for LBGs in \citet{2013ApJ...779...52J}. A central source being covered by patchy H {\sc i} gas clouds would lead to a small flux-averaged $N_{\rm HI}$ corresponding to a small $R^{\rm Ly\alpha}_{\rm IS}$. In this condition, Ly$\alpha$ photons could easily escape less affected by resonant scattering in the clouds. 
  \item The results of our kinematic analyses support the idea that the H {\sc i} column density is a key quantity determining Ly$\alpha$ emissivity. 

 \end{itemize}
 
In this kinematic study, we obtain $\Delta v_{\rm IS}$, $R_{\rm IS}^{\rm Ly\alpha}$, and $f_c$ only for objects with a moderate Ly$\alpha$ EW of $20-100$\,\AA\, which overlaps with the Ly$\alpha$ EW range of LBG samples in e.g., \citet{2003ApJ...588...65S}. We need to estimate these quantities for objects with a higher Ly$\alpha$ EW in order to check whether such objects follow the kinematic trends found in this study.

%% ACKNOWLEDGMENTS
%%%%%%%%%%%%%%%%%%%%%%%%%%%%%%%%%%%%%%%%%%%%%%%%
%%%%%%%%%%%%%%%%%%%%%%%%%%%%%%%%%%%%%%%%%%%%%%%%

\acknowledgments

We would like to thank Anne Verhamme, Zheng Zheng, Lennox L. Cowie, Esther M. Hu, and James E. Rhoads for useful discussion, and an anonymous referee, Mark Dijkstra, and Lucia Guaita for constructive comments. This paper is based on data collected with the Subaru Telescope, which is operated by the National Astronomical Observatory of Japan. Some of the data presented herein were obtained at the W.M. Keck Observatory, which is operated as a scientific partnership among the California Institute of Technology, the University of California and the National Aeronautics and Space Administration. The Observatory was made possible by the generous financial support of the W.M. Keck Foundation. The reduction pipeline used to reduce the LRIS data was developed at UC Berkeley with support from NSF grant AST-0071048. This work is based on observations taken by the CANDELS Multi-Cycle Treasury Program with the NASA/ESA {\it HST}, which is operated by the Association of Universities for Research in Astronomy, Inc., under NASA contract NAS5-26555. The NB387 data used in this work were collected at the Subaru Telescope, which is operated by the National Astronomical Observatory of Japan. This work is based in part on observations made with the {\it Spitzer Space Telescope}, which is operated by the Jet Propulsion Laboratory, California Institute of Technology under a contract with NASA. Support for this work was provided by NASA through an award issued by JPL/Caltech. This work was supported by World Premier International Research Center Initiative (WPI Initiative), MEXT, Japan. This work was supported by KAKENHI (23244025) and (21244013) Grant-in-Aid for Scientific Research (A) through Japan Society for the Promotion of Science (JSPS). MR was supported by NSF grant AST-1108815.

{\it Facilities:} \facility{Subaru (Suprime-Cam, FMOS)}, \facility{Keck:I (LRIS)}, \facility{Magellan:Baade (IMACS)}.

%% REFERENCE
%%%%%%%%%%%%%%%%%%%%%%%%%%%%%%%%%%%%%%%%%%%%%%%%
%%%%%%%%%%%%%%%%%%%%%%%%%%%%%%%%%%%%%%%%%%%%%%%%

\bibliographystyle{apj}
\bibliography{reference}

\end{document}